\documentclass[12pt]{article}
\pdfoutput=1
\usepackage{epsfig}
\usepackage[utf8]{inputenc}
\usepackage{epsfig}
\usepackage{amsfonts}
\usepackage{amssymb}
\usepackage{mathtools}
\usepackage{amsmath}
\usepackage{cancel}
\usepackage{color}
\usepackage{slashed}
\usepackage{cite}
\usepackage{calligra}
\usepackage{caption}
\usepackage{subcaption}
\usepackage{comment}
\usepackage{marginnote}
\usepackage{accents}
\usepackage{url}

\allowdisplaybreaks

\begin{document}

\begin{center}

\vspace{1cm}

\renewcommand{\thefootnote}{\fnsymbol{footnote}}

{\bf \Large Amplitudes in fishnet theories in diverse dimensions and box ladder diagrams} \vspace{1.0cm}

{\large L.V. Bork$^{1,2,3}$, R.M. Iakhibbaev$^{3}$, N.B. Muzhichkov$^{2,4}$, E.S. Sozinov$^{2,5}$}
\footnote{E-mail:
~\rm{bork@itep.ru},~\rm{ravmarat@gmail.com},~\rm{muzhnikita@gmail.com},~\rm{e.s.sozinov@gmail.com}}
\vspace{0.5cm}

{\it
$^1$Institute for Theoretical and Experimental Physics, Moscow, Russia,\\
$^2$The Center for Fundamental and Applied Research, \\ All-Russia
Research Institute of Automatics, Moscow, Russia, \\
$^3$Bogoliubov Laboratory of Theoretical Physics, \\ Joint
Institute for Nuclear Research, Dubna, Russia\\ 
$^4$Institute for Theoretical and Mathematical Physics, MSU, Moscow, Russia,\\
$^5$National Research Nuclear University MEPhI, Moscow, Russia
}

\renewcommand{\thefootnote}{\arabic{footnote}}
\setcounter{footnote}{0}

\vspace{1cm}

\abstract{We investigate properties of four-point colour ordered scattering amplitudes in $D=6$ fishnet CFT. We show that these amplitudes are related via a very simple relation to their $D=4$ counterparts previously considered in the literature. Exploiting this relation, we obtain a closed expression for these amplitudes and investigate its behaviour at weak and strong coupling. As a by product of this investigation, we also obtain a generating function for on-shell $D=6$ Box ladder diagrams with $l$ rungs.}
\end{center}

\begin{center}
\end{center}

\newpage

\tableofcontents{}\vspace{0.5cm}

\section{Introduction}\label{s1}
In the last decade, there has been substantial progress in understanding the structure of S-matrices (amplitudes) of gauge theories in various dimensions. $D=4$ $\mathcal{N}=4$ SYM is the most famous example. See \cite{Henrietta_Amplitudes,Talesof1001Gluons}\footnote{See also the list of talks at the \emph{Amplitudes} 2020 conference \url{https://indico.cern.ch/event/908370/timetable/}.} and reference therein for a review. This progress was initially achieved through to the extensive use of new analytical computational approaches and ideas such as generalised unitarity and spinor helicity/momentum twistor formalisms. In the case of $\mathcal{N}=4$ SYM, this allows one to uncover the integrable structure of the theory which in turn allows one to obtain even more analytical results about the structure of amplitudes and correlation functions, both approximate and exact with respect to the perturbation theory (PT). 

It was also recently discovered that there exists an integrable limit of $\mathcal{N}=4$ SYM which reduces the full $\mathcal{N}=4$ SYM theory to the theory of a pair of interacting adjoint scalar $SU(N_c)$ fields \cite{Fokken:2013aea,Sieg:2016vap,Grabner:2017pgm,Gurdogan:2015csr}. This theory is much simpler than the full $\mathcal{N}=4$ SYM but still remains non-trivial enough to be an interesting object for investigation. The Feynman diagrams that arise in this theory have in fact been known for quite some time under the nickname "fishnet graphs" \cite{ZamolodchikovFishnet}, and so this limit of $\mathcal{N}=4$ SYM is usually called the fishnet theory in the literature. Within this direction of investigation a plethora of interesting results was obtained \cite{ExactCorrelationFunctionsFishnet,FishnetGeneralD,Gromov:2017cja,Kazakov:2019laa,Basso:2019xay,Karananas:2019fox,Kazakov:2018gcy,Derkachov:2018rot,Kazakov:2018ugh,Chicherin:2017cns,Chicherin:2017frs,Loebbert:2020glj,Loebbert:2020tje,Loebbert:2020hxk,Loebbert:2019vcj,Adamo:2019lor,FishnetSpinChain,KorchFishNet,Chowdhury:2019hns,Chowdhury:2020tbn,Pittelli:2019ceq,Kazakov:2018gcy,Levkovich-Maslyuk:2020rlp} which range from new computation methods and analytical expressions for multiloop Feynman integrals \cite{Loebbert:2019vcj,Loebbert:2020glj} to a new all-loop representation for the scattering amplitudes \cite{KorchFishNet}.

It was also discovered that in fact there exists a whole family of fishnet theories that can be defined in arbitrary dimension $D$ \cite{FishnetGeneralD}. It is unknown whether a parent theory exists for this type of models beyond $D=4$. However, the $D=6$ case provides some interesting hints.

In this article, we are going to investigate the properties of amplitudes in some particular case of the $D=6$ fishnet model. Our main focus will be on the colour ordered four-point amplitudes in this theory. 
Our motivation for such an investigation was initially given by the results of \cite{Lip} and also \cite{BKIT}, where the properties of the four-point amplitude in the hypothetical
$D=6$ $SU(N_c)$ planar gauge theory with dual conformal symmetry were discussed. Also, in the same article  \cite{Lip} some speculations were presented regarding the possibility that such a hypothetical theory may be none other than the mysterious $(2,0)$ SYM. We found out that the $D=6$ scalar integrals discussed in \cite{Lip} as well as some relations between $D=4$ and $D=6$ scalar integrals reappear in our $D=6$ fishnet theory. 
We also found out that our case of the $D=6$ fishnet theory (four-point amplitude in this theory should be precise) is directly related to the $D=4$ counterpart, so the results obtained in this article are relevant to the $D=4$ fishnets as well.

Our article is organised as follows. In Section 2, we give a brief introduction to the fishnet models in arbitrary dimension. We comment on the integrability properties of these models and typical observables, which one computes in such theories. We also outline the special case of the $D=6$ model with the deformation parameter $\omega$ equal to 1, which will be our prime interest in this article.

In Section 3, we show how the sub-sector of the $D=6$ model with $\omega=1$ is related to the $D=4$ case so that at the level of double-trace four-point colour ordered amplitudes these models are essentially identical. As a by product of this consideration, we also obtain the generating function for the $D=6$ four-point on-shell Box integrals with $l$-rungs. This result is interesting by itself because the $D=6$ box integrals appear in various contexts \cite{Dennen:2010dh,PanzerHyperInt,BorkD6}, including phenomenological ones \cite{Bern:2002tk}. 

In Section 4, we investigate the weak coupling behaviour of the four-point amplitude in the $D=6$, $\omega=1$ model. We obtain the integral representation for the generating function of the $D=6$ Box integrals in terms of the Lerch transcendent Zeta functions. We show that one can, in principle, use this integral representation evaluated
at a fixed kinematic point to bootstrap perturbative expansion for the $l$-loop Box functions or equivalently for the four-point amplitude for arbitrary kinematics. We also discuss the Regge behaviour of the $D=6$ Box integrals and give some all loop predictions for the logarithmically enhanced terms in these integrals.

In Section 5, using the integral representation of the Box generating function obtained in the previous section, we consider the strong coupling limit of the four-point amplitudes in the $D=4$ and $D=6$ fishnet models.

In Section 6, we make short comments on the relation of the results obtained here with some other known $D=6$ models.

In conclusion, we give a brief summary of the results obtained in this article. In Appendices, we list the answers for one; two- and three-loop $D=6$ Box integrals and their asymptotic behaviour as well as some supplementary details regarding the strong coupling limit computations. In addition, we also make some comments regarding the structure of the fixed point of the $D$=6, $\omega=1$ fishnet theory considered here.

\section{Fishnet models in arbitrary dimension}\label{s2}
\begin{figure}[ht]
 \begin{center}
  \epsfxsize=10cm
 \epsffile{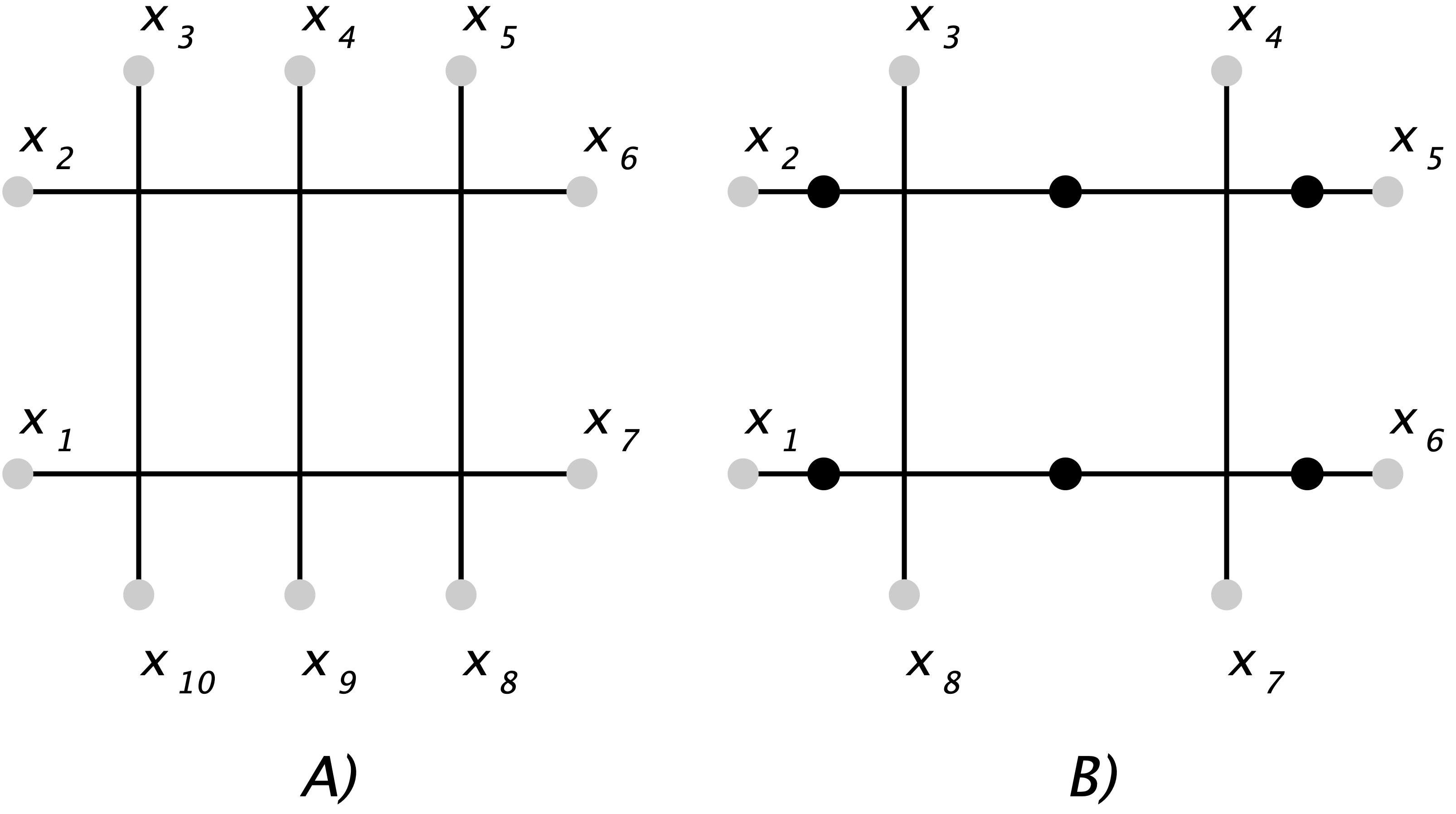}
 \end{center}\vspace{-0.2cm}
 \caption{Examples of fishnet graphs. A) graph corresponds to the only one contribution to the single-trace 10-point correlation function $\langle\mbox{tr}(\phi_{1}(x_1)\phi_{1}(x_2)\ldots\phi_{2}(x_{10}))\rangle$ in the $D=4$, $\omega=1$ model. The propagators of the $\phi_1$ and $\phi_2$ fields are identical in this case. B) graph corresponds to the only contribution to the single-trace 8-point correlation function $\langle\mbox{tr}(\chi(x_1)\chi(x_2)\ldots\phi(x_{8}))\rangle$ in the $D=6$, $\omega=1$ model. Dots on the lines correspond to the non-standard power of the propagator of the $\chi$ field ($1/x^2$ in coordinate space, which translates to $1/p^4$  in momentum space).}\label{fig1}
 \end{figure}
In arbitrary dimension $D$ one can consider the theory of a pair of interacting scalar fields $(\phi_1,\phi_2)$  given by the following Lagrangian \cite{FishnetGeneralD}:
\begin{eqnarray}\label{Lmain}
\mathcal{L}_{main}=N_c~\mbox{tr}\left(~\phi_1^*(\partial^2)^{\omega}\phi_1+\phi_2^*(\partial^2)^{D/2-\omega}\phi_2+
(4\pi)^{D/2}g^2~\phi_1^*\phi_2^*\phi_1\phi_2~\right),
\end{eqnarray} 
where $\phi_i$ transforms under the adjoint representation of $SU(N_c)$ and parameter $\omega\in (0,\frac{D}{2})$; $g$ is the dimensionless (for arbitrary $D$) coupling constant and large $N_c$ limit is implemented. For the general $D$ and $\omega$ this theory is non-local and also non-unitary. If this Lagrangian is accompanied by double-trace terms of the form \cite{FishnetGeneralD,Grabner:2017pgm,Pomoni:2008de,Dymarsky:2005uh}
\begin{eqnarray}\label{Ldtrace1}
\mathcal{L}_{c.t.}^{(D,\omega)}=~\sum_i \alpha_i(g)\mbox{tr}\left(\mathcal{O}_2^{(i)}\right)\mbox{tr}\left(\tilde{\mathcal{O}}_2^{(i)}\right),
\end{eqnarray}
where $\mathcal{O}_2^{(i)}$ and $\tilde{\mathcal{O}}_2^{(i)}$ are some quadratic monomials of the $(\phi_1,\phi_2)$ fields, the theory given by
\begin{eqnarray}\label{Lfull}
\mathcal{L}=\mathcal{L}_{main}+\mathcal{L}_{c.t.}^{(D,\omega)},
\end{eqnarray}
is conformal invariant (for an appropriate choice of $\alpha_i(g)$ \cite{Pomoni:2008de,Dymarsky:2005uh}) and integrable in a sense that the whole spectrum of anomalous dimensions of local operators is known and all correlation functions of these
operators and also scattering amplitudes can be in principle evaluated in a closed form. 

This is, roughly speaking, the consequence of the fact that in a theory like this in the large $N_c$ limit all Feynman diagrams, which contribute to the aforementioned observables, have the iterative structure and all can be represented in coordinate space as a consecutive application of the so called "graph building" operator \cite{ExactCorrelationFunctionsFishnet,FishnetGeneralD} (see fig.\ref{fig2}):
\begin{figure}[t]
 \begin{center}
  \epsfxsize=10cm
 \epsffile{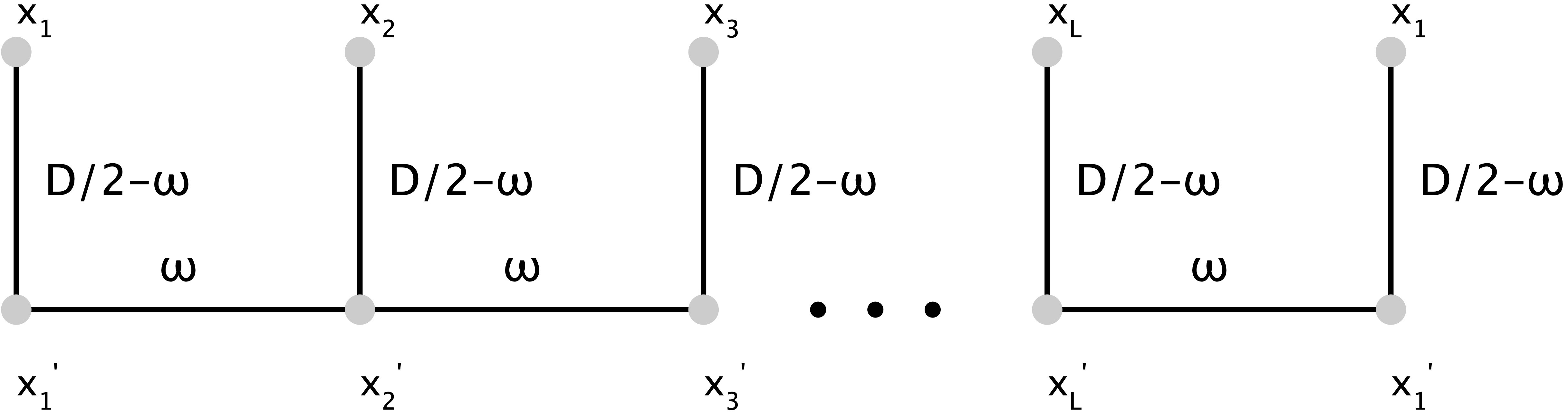}
 \end{center}\vspace{-0.2cm}
 \caption{Graphical representation of $\mathcal{H}_L$. The powers of the propagators in coordinate space are explicitly written. The periodical boundary condition is implemented for the double-trace case.}\label{fig2}
 \end{figure}
\begin{equation}\label{Hoper}
\begin{gathered}
\Phi^{(l)}(x_1,\ldots,x_L)=\mathcal{H}_L\Phi^{(l-2)}(x_1,\ldots,x_L)=\\
=\frac{1}{\pi^{DL/2}}\int \frac{d^Dx_{1'}\ldots d^Dx_{L'}~\Phi^{(l')}(x_{1'},\ldots,x_{L'})}
{|x_{11'}|^{D-2\omega}\ldots|x_{LL'}|^{D-2\omega}|x_{1'2'}|^{2\omega}\ldots|x_{L'1'}|^{2\omega}},
\end{gathered}
\end{equation}
where $\Phi^{(l)}(x_1,\ldots,x_L)$ is the $l$-loop Feynman graph. See fig.\ref{fig1} A) for the $D=4,\omega=1$ example; the
$\mathcal{H}_L$ operator itself is the Hamiltonian of the deformed $SO(1,1+D)$ Heisenberg spin chain of length $L$ in the non-compact representation \cite{FishnetGeneralD}. These models are known to be integrable. The spectrum and eigenfunctions of such Hamiltonians are known at least for some values of $D$ and $\omega$ \cite{FishnetSpinChain,IsaevStarTriangle}. One can use 
this information to obtain closed expressions for the correlation functions and scattering amplitudes
\cite{ExactCorrelationFunctionsFishnet,FishnetGeneralD,KorchFishNet}.

As an illustrative example, let us consider the $D=4,\omega=1$ case in more detail \cite{Grabner:2017pgm,KorchFishNet}. In this case, the theory is local. It is also conformal invariant but non-unitary \cite{Grabner:2017pgm} due to the "wrong" structure of the interaction term in (\ref{Lmain}). The explicit form of $\mathcal{L}_{c.t.}$ is
\begin{eqnarray}\label{DTraceD4}
\mathcal{L}_{c.t.}^{(4,1)}&=& (4\pi)^2\alpha_1(g)\left(\mbox{tr}\left(\phi_1^*\phi_1^*\right)\mbox{tr}\left(\phi_1\phi_1\right)+
\mbox{tr}\left(\phi_2^*\phi_2^*\right)\mbox{tr}\left(\phi_2\phi_2\right)\right)\nonumber\\
&+&(4\pi)^2\alpha_2(g)\left(\mbox{tr}\left(\phi_1^*\phi_2^*\right)\mbox{tr}\left(\phi_1\phi_2\right)+
\mbox{tr}\left(\phi_1\phi_2^*\right)\mbox{tr}\left(\phi_1^*\phi_2\right)\right),
\end{eqnarray}
where $\alpha_1(g)$ and $\alpha_2(g)$ are given by \cite{Grabner:2017pgm,KorchFishNet}
\begin{eqnarray}
\alpha_1(g)=\frac{ig^2}{2}+O(g^4),~\alpha_2(g)=g^2.
\end{eqnarray}
The single-trace correlation functions such as
\begin{eqnarray}
\langle\mbox{tr}\left(\phi_1(x_1)\phi_1(x_2)\ldots\phi_2(x_{9})\phi_2(x_{10})\right)\rangle
\end{eqnarray}
are protected from quantum corrections and are given by the single fishnet Feynman diagram to all orders of PT (see fig.\ref{fig1}) \cite{ExactCorrelationFunctionsFishnet}. Quantum corrections appear at the level of multi-trace correlation functions and we have non-trivial infinite PT series \cite{ExactCorrelationFunctionsFishnet}. For example, three-loop corrections to the four-point correlation function 
\begin{figure}[t]
 \begin{center}
  \epsfxsize=14cm
 \epsffile{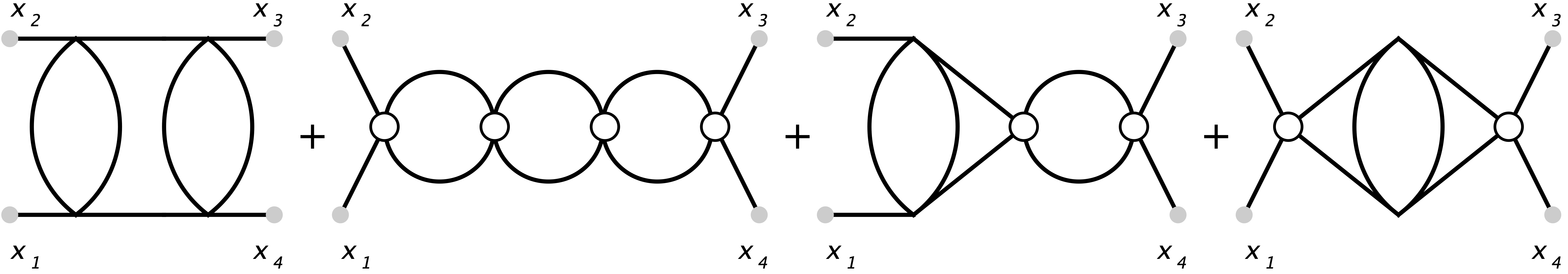}
 \end{center}\vspace{-0.2cm}
 \caption{Example of s-channel three loop diagrams in coordinate space which contribute to the double-trace correlation 
 function $\langle \mbox{tr}\left(\phi_1(x_1)\phi_1(x_2)\right) 
 \mbox{tr}\left(\phi_1^*(x_3)\phi_1^*(x_4)\right)\rangle$ in the $D=4$, $\omega=1$ fishnet theory. White blobs correspond to the double-trace interactions. The full contribution is given by adding the same diagrams with $x_1\leftrightarrow x_2$.}\label{fig3}
 \end{figure}
\begin{eqnarray}
G_4(x_1,\ldots,x_4)=\langle \mbox{tr}\left(\phi_1(x_1)\phi_1(x_2)\right) 
\mbox{tr}\left(\phi_1^*(x_3)\phi_1^*(x_4)\right)\rangle
\end{eqnarray} 
are presented in fig.\ref{fig3}. One can see that all these diagrams have the iterative structure\footnote{The double-trace interaction contributions can be associated with the action of another operator $\mathcal{V}_2$ with a known spectrum. However, such contributions are cancelled out in the final result\cite{ExactCorrelationFunctionsFishnet,FishnetGeneralD}.}. Using this fact and the knowledge of $\mathcal{H}_L$ eigenvalues/eigenfunctions of the $SO(1,5)$ non-compact spin chain of length $L=2$, one can obtain a closed expression for such a correlation function that is valid in all orders of PT \cite{ExactCorrelationFunctionsFishnet}:
\begin{eqnarray}\label{4PointCorrExact}
G_4(x_1,\ldots,x_4)=\sum_{J=0}^{\infty}\int_{-\infty}^{+\infty}d\nu\frac{\mu(\nu,J)}{h(\nu,J)-g^4}\Pi_{\nu,J}(x_1,\ldots,x_4)+(x_1\leftrightarrow x_2),
\end{eqnarray}
here
\begin{eqnarray}
\mu(\nu,J)&=&16\pi\frac{\nu^2(4\nu^2+(J+1)^2)(J+1)}{2^J},\\
h(\mu,J)&=&\left(\nu^2 + \frac{J^2}{4}\right) \left(\nu^2 + \frac{(J + 2)^2}{4}\right),
\end{eqnarray}
while $\Pi_{\nu,J}$ is the known function of the $\nu,J$ and coordinates whose explicit form is not important for us now. It can be found in \cite{ExactCorrelationFunctionsFishnet}. Here $\nu$ and $J$ are parameters of the irreducible principal series representation of the conformal group \cite{ExactCorrelationFunctionsFishnet,FishnetGeneralD}, $\nu$ is real and $J$ is integer and non-negative. It should also be mentioned that $h(\mu,J)^{-1}$ is the eigenvalue of the $\mathcal{H}_2$ operator and that the equation $h(\mu,J)-g^4=0$ defines the spectrum of anomalous dimensions of local operators in the theory of form $\mbox{tr}(\phi_1\partial^J\phi_1(\phi_2^*\phi_2)^k)$.
A similar representation of four-point correlation functions of other operators can be obtained as well \cite{ExactCorrelationFunctionsFishnet}. It should also be mentioned that since the $D=4,\omega=1$ fishnet model is an example of CFT, the non-trivial dependence of $\Pi_{\nu,J}$ on coordinates is condensed in a pair of conformal cross ratios in full agreement with general CFT expectations. It is also important to note that the dependence on the $\alpha_i(g)$ factors from the double-trace terms in the Lagrangian, which we know in general only in the form of PT series, is cancelled out in (\ref{4PointCorrExact}), so the dependence on the coupling constant $g$ in (\ref{4PointCorrExact}) is known for arbitrary values of $g$. 

The knowledge of the correlation function in a closed form of (\ref{4PointCorrExact}) allows one to obtain similar closed expressions for the scattering amplitudes using the Lehmann-Symanzik-Zimmerman reduction formula (LSZ) \cite{KorchFishNet}. More accurately, one can perform colour decomposition and separate leading single-trace and suppressed multi-trace contributions to the amplitudes. At the level of four-point amplitudes, this results in the following colour structure of the amplitude
\begin{eqnarray}
\mathcal{A}_4^{D=4}\sim\sum_{perm.}\left( N_c\mbox{tr}(T^{a_1}T^{a_2}T^{a_3}T^{a_4})A^{D=4,sing.}_4+
\mbox{tr}(T^{a_1}T^{a_2})\mbox{tr}(T^{a_3}T^{a_4})A^{D=4,doub.}_4\right),
\end{eqnarray}
where $\mathcal{A}_4$ is the physical amplitude, $A^{D=4,sing.}_4$ is the single-trace colour ordered partial amplitude and $A^{D=4,doub.}_4$ is the double-trace one. Hereafter we will omit these superscripts and will consider the $A^{D,doub.}_4$ amplitudes only.
In fishnet theories, all single-trace partial amplitudes are protected from quantum corrections and in the four-particle case are just equal to constant. The partial amplitude $A_{4,\phi_1\phi_2\phi_1^*\phi_2^*}^{D=4}$, which "mixes" $\phi_1$ and $\phi_2$ particles, is also protected at the double-trace level. The double-trace amplitudes $A_{4,\phi_1\phi_1\phi_1^*\phi_1^*}^{D=4}$ and $A_{\phi_2\phi_2\phi_2^*\phi_2^*}^{D=4}$ are not protected and are non-trivial. They are equal to each other and hereafter we drop the particle content subscript to simplify the notation as well. One can use representation (\ref{4PointCorrExact}) to obtain closed expressions for these amplitudes via LSZ. The result is given by (hereafter the delta function of total momentum conservation will also be omitted) \cite{KorchFishNet}:
\begin{equation}\label{Ampld4}
A_4^{D=4}=A_4^{D=4,u}+A_4^{D=4,t}=\int_{-\infty}^{+\infty} d\nu\sum_{J \geq 0}^{\infty}\frac{\mu(\nu,J)}{h(\nu,J)-g^4}\Omega_{\nu,J}(z)+(z\rightarrow -z),
\end{equation}
where the $\Omega_{\nu,J}(z)$ function is given by
\begin{equation}\label{Ampld4Omega}
\Omega_{\nu,J}(z)=\frac{2^J}{\pi^2}\sinh^2(\pi\nu+i\pi J/2)\sum_{k=0}^J\frac{P_k(z)P_{J-k}(z)}{(J/2-k)^2+\nu^2},
\end{equation}
with $z=1-2u/s$; $s,t,u$ are the standard Mandelstam variables related as $s+t+u=0$; $P_n(z)$ is the Legendre polynomial. Here we explicitly indicate the contributions from the $t$ and $u$ channel diagrams as $A_4^{D=4,t}$ and $A_4^{D=4,u}$, respectively. Note that the contribution from each channel can be written in a closed form separately.  
Using this expression as a starting point, one can investigate the behaviour of the amplitude in various limits for, in principle, an arbitrary value of the coupling constant.

In this article, we are going, however, to concentrate our attention on another example of the fishnet theory, namely on the $D=6,\omega=1$ theory. This theory is also conformal and local but now we have two different types of kinetic terms in the Lagrangian: with the $1/p^2$ and $1/p^4$ propagators in momentum space. We will label the fields with the $1/p^2$ and $1/p^4$ propagators $\phi_1\equiv\phi$ and $\phi_2\equiv\chi$, respectively. The single-trace Lagrangian of the theory is then given by
\begin{eqnarray}\label{D6LagrMain}
\mathcal{L}_{main}=N_c~\mbox{tr}\left(~\phi^*(\partial^2)\phi+\chi^*(\partial^4)\chi+
(4\pi)^3g^2~\phi^*\chi^*\phi\chi~\right).
\end{eqnarray} 
At the level of single-trace correlation functions, the behaviour of this theory is expected to be similar to the $D=4$ case. The single-trace correlation functions are given by the single Feynman diagram of a fishnet type and are protected from quantum corrections. For example, the correlation function 
\begin{eqnarray}
\langle\mbox{tr}\left(\chi(x_1)\chi(x_2)\ldots\phi(x_7)\phi(x_8)\right)\rangle
\end{eqnarray}
is given by the Feynman diagram represented in fig.\ref{fig1} B). The Feynman diagrams of this type were first discussed in \cite{Lip} in the context of the scattering amplitudes in the hypothetical $D=6$ gauge theory with dual conformal symmetry. 
Our main focus of investigation, as was claimed before, will be not on correlation functions themselves but scattering amplitudes. Namely we will focus on the four-point colour ordered amplitudes. Inspired by the results of \cite{Lip,KorchFishNet}, we expect that these amplitudes will have a relatively simple and interesting structure. 

Let us now discuss the structure of the four-point scattering amplitudes in this theory in more detail. Similar to the $D=4$ case, single-trace amplitudes are protected from quantum corrections and are kinematically independent (equal to constant). Also, similarly to the $D=4$ case, the four-point amplitudes factorise into the $\phi$ and $\chi$ subsectors (amplitudes with all $\phi$ external legs or with only $\chi$ external legs) with no mixing between them. In contrast to the $D=4$ case, the amplitudes in these subsectors behaves rather differently. We will first focus on the $\phi$ subsector and make some comments regarding the structure of the $\chi$-subsector in the appendix. 

\begin{figure}[t]
 \begin{center}
  \epsfxsize=13cm
 \epsffile{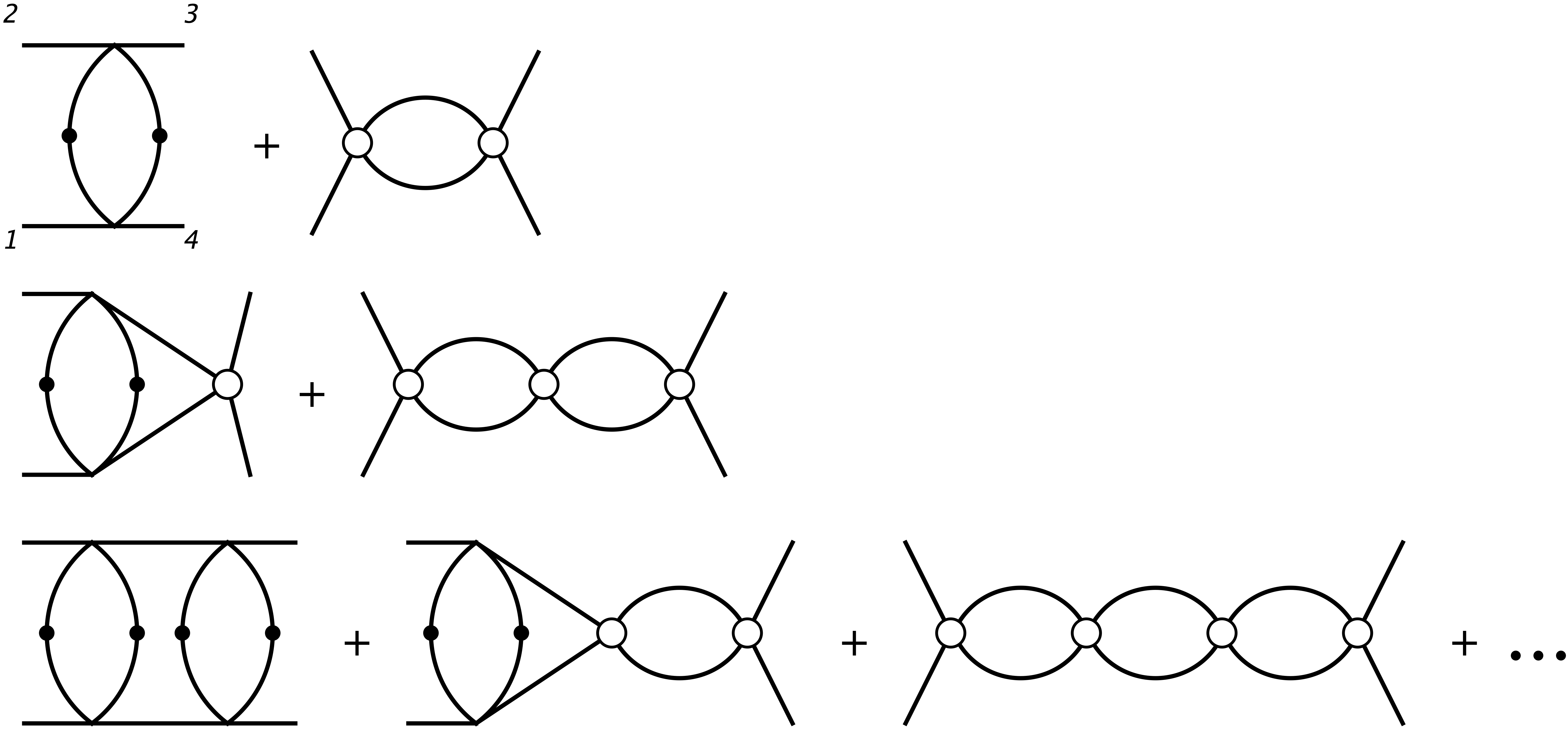}
 \end{center}\vspace{-0.2cm}
 \caption{Feynman diagrams contributing to the first orders of perturbative expansion of the $A_4^{D=6}$ amplitude in the $\phi$ sector in the $t$ channel. White blobs correspond to double-trace interactions, which in this particular case are equal to zero, so only the first diagram on the first and third rows gives non-trivial contributions.}\label{fig4}
 \end{figure}

The contribution to the scattering amplitudes of the $\phi$ sector from the single-trace Lagrangian in the weak coupling limit is given by the
ladder type Feynman diagrams represented in fig.\ref{fig4}. Hereafter we will label this type of amplitude as 
$A_{4,\phi_1\phi_1\phi_1^*\phi_1^*}^{D=6}\equiv A_4^{D=6}$. After a short investigation one can see that this diagram at $l$-loops is actually UV finite, so no double-trace terms in the Lagrangian are required in this sector. To evaluate the four-point amplitude, one can use the general approach of \cite{KorchFishNet,FishnetGeneralD} based on the graph building operator and integrability of the $SO(1,D+1)$ non-compact spin chain of length $L=2$ with $D=6$, $\omega=1$.
In the next section, we will obtain a closed expression for $ A_4^{D=6}$ and show that the scattering amplitudes at the $\phi$ sectors in $D=6$ and $D=4$ are connected via a simple relation.

As for the $\chi$ sector scattering amplitudes in the weak coupling limit, the situation is different and it is likely that their UV perturbative finiteness will require double-trace terms of the $\chi$ field interaction:
\begin{eqnarray}\label{D6counterterms}
\mathcal{L}_{c.t.}^{(6,1)}/(4\pi)^3=g\left(
\mbox{tr}\left(\phi\chi\right)\mbox{tr}\left(\phi^*\chi^*\right)
+\mbox{tr}\left(\phi^*\chi\right)\mbox{tr}\left(\phi\chi^*\right)
\right)+\mbox{possible}~\chi~\mbox{interactions}.
\end{eqnarray}
\begin{figure}[t]
 \begin{center}
  \epsfxsize=7cm
 \epsffile{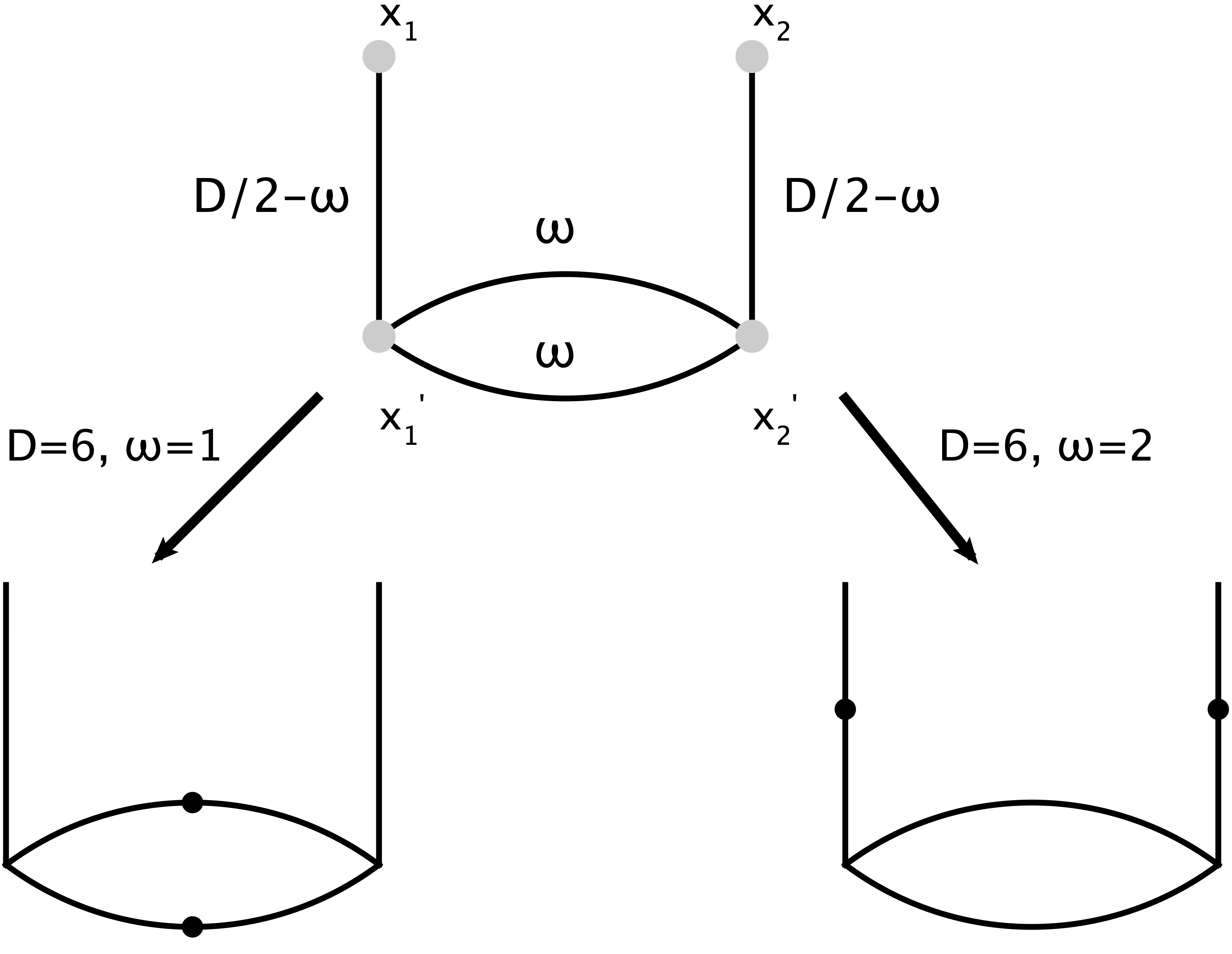}
 \end{center}\vspace{-0.2cm}
 \caption{Graphical representation of the Graph building operator $\mathcal{H}_2$ and the corresponding Feynman diagram element whose $\mathcal{H}_2$ will generate in momentum space in the $D=6$, $\omega=1$ and $\omega=2$ cases.}\label{figGBOperL2}
 \end{figure}
We will discuss these questions in Appendix \ref{a4}. 
To compute such an amplitude, one can possibly apply the graph building operator approach. 
However, now one has to consider the $SO(1,D+1)$ non-compact spin chain with the parameters $L=2$, $D=6$ and $\omega=2$. See fig.\ref{figGBOperL2}.

\section{$D=4$ - $D=6$ correspondence and $A_4$ amplitude in Fishnet models as a generating function of $D=6$ Box ladder diagrams}\label{s3}
As was mentioned before, one can expect that for the amplitudes in the $D=6$ fishnet theory in the $\phi$ sector, one can obtain a closed expression similar to (\ref{Ampld4}) using the same steps as in \cite{KorchFishNet}. This is, in principle, true but before doing so, one can make the following observation.

\begin{figure}[t]
 \begin{center}
  \epsfxsize=12cm
 \epsffile{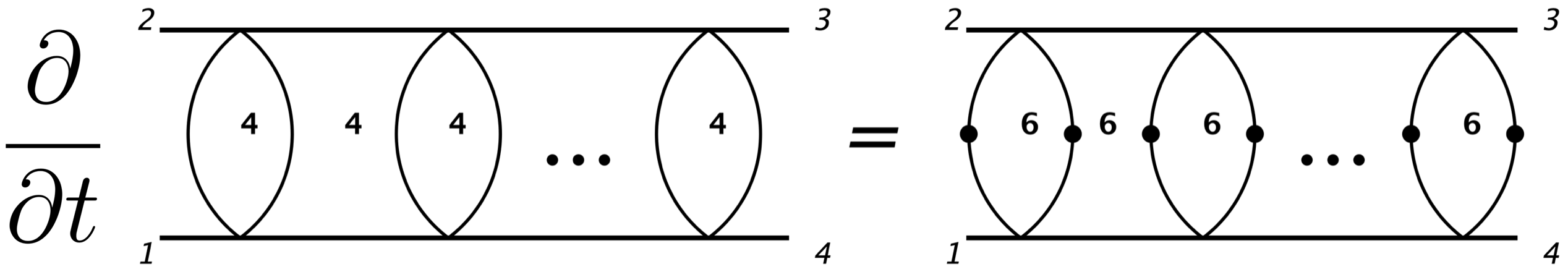}
 \end{center}\vspace{-0.2cm}
 \caption{Relation between the $D=4$ and $D=6$ ladder diagrams. Numbers in the diagram correspond to the dimensionality of loop integration.}\label{figLoopRelation}
 \end{figure}

There exists a relation which connects the Feynman integrals in $D$ dimension with the integral with the same topology in $D+2$ dimension but with the propagators raised in higher powers ("dots on the lines") \cite{Smirnov_book_2}. For four-point integrals with massless lines and box-like topology
and on-shell external legs this relation takes the form:
\begin{equation}\label{ddp2relation}
\frac{\partial}{\partial t}I_4(s,t;\ldots;\alpha_{v_1},\ldots,\alpha_{v_n};D)=
I_4(s,t;\ldots;\alpha_{v_1}+1,\ldots,\alpha_{v_n}+1;D+2)
\end{equation}
where $I_4$ is the four-point Feynman integral written in the $\alpha$-representation \cite{Smirnov_book_2}:
\begin{equation}\label{alpha1}
I_4(s,t;\ldots;\alpha_{v_1},\ldots,\alpha_{v_n};D)\sim\int^\infty_0 \prod^{n+m}_{i=1} d \alpha_i \; \prod^{n+m}_{i=1}\alpha^{\nu_i-1}_i  \frac{e^{i (\mathcal{V}_s s+\mathcal{V}_t t)/{\mathcal{U}}}}{\mathcal{U}^{d/2}}.
\end{equation}
Here $n$ is the number of vertical rungs and $m$ is the number of horizontal rungs; the $\alpha$-parameters
associated with the vertical rungs are labelled as $\alpha_{v_i}$; $\nu_i$ are the powers of propagators. For example, for fig.\ref{figLoopRelation} all $\nu_i=1$ for the $D=4$ diagram depicted here; $\mathcal{V}_s$, $\mathcal{V}_t$ and $\mathcal{U}$ are polynomials in the $\alpha$ parameters and in the case of box-like four-point topology 
$\mathcal{V}_t=\prod_i^n\alpha_{v_i}$, which makes relation (\ref{ddp2relation}) obvious.

\begin{figure}[t]
 \begin{center}
  \epsfxsize=12cm
 \epsffile{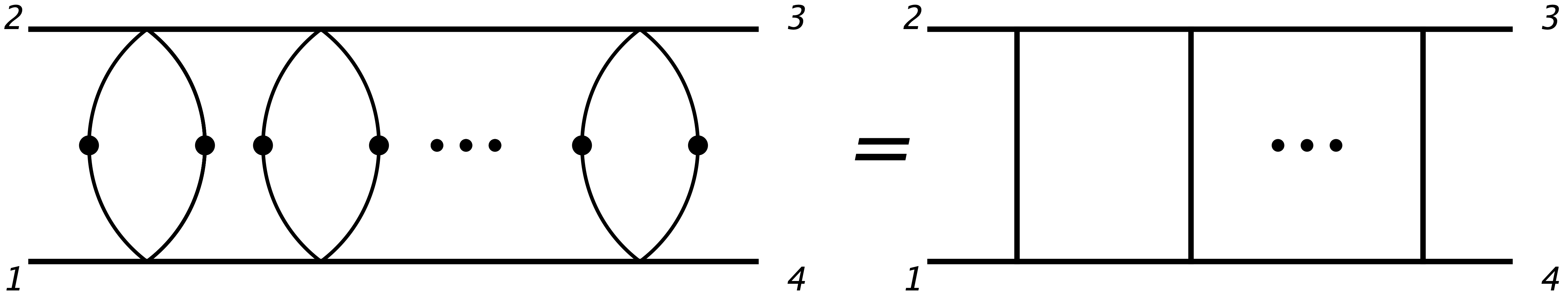}
 \end{center}\vspace{-0.2cm}
 \caption{Equivalent forms of the $D=6$ on-shell Box ladder diagram.}\label{fig5}
 \end{figure}

Noting that contributions from the double-trace terms in the $D=4$ Lagrangian (\ref{DTraceD4}) depend only on $s/\mu^2$ and taking a derivative of $A_4^{D=4}$ with respect to $z$, we obtain ($\partial z \sim \partial t$):
\begin{equation}\label{AmplD6ClosedForm}
A_4^{D=6}(s,z,g)=\frac{1}{s}\frac{\partial}{\partial z} A_4^{D=4}(z,g)=\frac{1}{s}\int_{-\infty}^{+\infty} d\nu\sum_{J \geq 0}^{\infty}\frac{\mu(\nu,J)}{h(\nu,J)-g^4}\frac{\partial\Omega_{\nu,J}(z)}{\partial z}+(z\rightarrow -z).
\end{equation}
In addition, one can note that the bubble subgraphs with the $1/p^4$ propogators in $D=6$ can be exactly evaluated back to $1/p^2$:
\begin{equation}\label{alpha1}
\int \frac{d^6k}{k^4(k-Q)^4}\sim\frac{1}{Q^2}
\end{equation}
so that the diagrams contributing to $A_4^{D=6}$ are actually just the $D=6$ Box ladder diagrams $B^{(l)}(s,t)$ (see fig.\ref{fig5} and Appendix \ref{a1} for details). We have verified these relations by directly evaluating the box integrals up to 4 loops and comparing the obtained results with the derivatives of perturbative calculations of \cite{KorchFishNet}. Note also that one, two and three loop $D=6$ results are equivalent to the derivatives of three, five and seven loop results in the $D=4$ theory. We have also rederived relation (\ref{AmplD6ClosedForm}) by applying the graph building operator approach in the large $z$ limit. 

The result (\ref{AmplD6ClosedForm}) is also interesting from the following point of view. The Lagrangian of the $D=6$ theory (\ref{D6LagrMain}) contains the $\chi$ fields with the $1/p^4$ kinetic term. Such Lagrangians are usually considered as pathological due to the common belief that higher derivatives generate ghost states with negative norm in the spectrum of the theory. Here at the four-point amplitude level we see that we have a closed sub-sector where all $1/p^4$ propagators cancel out in all orders of PT. Such a sector, however, will still  be formally non-unitary due to the "wrong" structure of the interaction term similar to the $D=4$ case. But no additional problems with ghost states etc. will arise.

Also, let us make the following observation. The $D=4$ and $D=6$ amplitudes (\ref{Ampld4}), (\ref{AmplD6ClosedForm}) are also IR finite, which is a counter example of the popular belief that no S-matrix exists in CFT's in the strict sense, i.e., without the introduction of some symmetry braking IR regulator due to the presence of massless states (particles) in the spectrum of the theory. In the $D=4$ case IR divergences appear in higher orders of $1/N_c$ expansion, but regardless of that double-trace four-point amplitudes in the $D=4$ and $D=6$ fishnet theories give an example of IR and UV finite amplitudes in CFT's. In the light of \cite{ConformalAnomalyFormFactors}, there is an open question whether such amplitudes preserve full conformal symmetry. However, in a naive sense, they certainly do because they are finite functions of the two dimensionless variables $z=1-2u/s$ and $g$ (there is also a trivial overall factor $1/s$ in the $D=6$ case). 

Summarizing, we have obtained the following:
\begin{itemize}
\item The four-point amplitude in the $D=6$, $\omega=1$ fishnet theory in the $\phi$ sector can be evaluated in a closed form and is given by (\ref{AmplD6ClosedForm}).

\item This amplitude ($t$ or $u$ channel contribution to this amplitude to be exact) can be considered as a generating function of the $D=6$ Box ladder diagrams $B^{(l)}(s,t)$: 
\begin{eqnarray}
A_4^{D=6}(s,z,g)&=&A_4^{D=6,u}(s,z,g)+A_4^{D=6,t}(s,-z,g),\nonumber\\
A_4^{D=6,u}(s,z,g)&=&\sum_{l=0}^{\infty}(g^4)^{l+1}~B^{(l)}(s,u).
\end{eqnarray}
\end{itemize}
In the next section we are going to apply these results to 
obtain information about the weak coupling limit of (\ref{AmplD6ClosedForm}).

\section{$A_4$ amplitude at weak coupling}\label{s4}
As a starting point, let us briefly discuss the weak coupling expansion of $A_4^{D=6}$ based on the Feynman diagrams. See the first and third row in fig.\ref{fig4}. It is convenient to use the dimensionless function $\mathcal{B}^{(l)}(x)$, $x=(z-1)/2=u/s$ instead of $B^{(l)}(s,u)$ which are related as $B^{(l)}=\mathcal{B}^{(l)}/s$, see Appendix \ref{a1} for details. In these notations we have: 
\begin{eqnarray}
A_4^{D=6,u}(s,z,g)=\frac{1}{s}\left(g^4~\mathcal{B}^{(0)}(x)+g^8~\mathcal{B}^{(1)}(x)+\ldots\right),
\end{eqnarray}
with:
\begin{eqnarray}
\mathcal{B}^{(0)}(x)=\frac{1}{x},~\mathcal{B}^{(1)}(x)=\frac{\log(x)^2+\pi^2}{2(1+x)}.
\end{eqnarray}
One can easily verify that these answers can be obtained by taking $\partial/\partial t$ from the corresponding
$D=4$ answers, which can be found in \cite{KorchFishNet}. Explicit answers for higher orders of PT up to four loops can be found in Appendix \ref{a1}. They all can be written in terms of the Harmonic Polylogarithms (HPLs) $H_{a_1,a_2,\ldots}(x)$
\cite{HPL_init}. In fact, one can show that this will be true for an arbitrary order of PT \cite{PanzerHyperInt}. We expect that at $l$ loops the answer for $~B^{(l)}$ can be written as a linear combination of $H_{a_1,\ldots,a_w}(x)$ with $a_i\in\{0,-1\}$ and $w \leq 2l$ \cite{PanzerHyperInt}.

The natural question to ask is what information about PT series we can extract from expression (\ref{AmplD6ClosedForm}). There can be several approaches \cite{KorchFishNet} to investigation of the weak coupling expansion of (\ref{AmplD6ClosedForm}). The first is to interchange $\nu$ integration and $J$ sum, and evaluate $\nu$ integral by residues. This will lead us to \cite{KorchFishNet}
\begin{eqnarray}\label{AJseriesRep}
A^{D=6,u}(s,z,g)=\frac{1}{s}\sum_{J=0}^{\infty}\sum_{i=1}^2\frac{2^{1-J}(J+1)((J+1)^2+4\nu^2_i)\nu_i}{(J+1)^2+4\nu_i+1}
\frac{\partial\Omega_{J,\nu_i}(z)}{\partial z},
\end{eqnarray}
where
\begin{eqnarray}\label{nu13}
\nu_1&=&-\frac{1}{2}\sqrt{-2-2J-J^2-2\sqrt{1+4g^4+2J+J^2}},\nonumber\\
\nu_2&=&-\frac{1}{2}\sqrt{-2-2J-J^2+2\sqrt{1+4g^4+2J+J^2}}.
\end{eqnarray}
The sum over $J$ in (\ref{AJseriesRep}), however, is expected to be divergent, at least for general value of $z$ due to the
exponential growth of $\Omega_{J,\nu_i}(z)$ as a function of $\nu$ for fixed $J$ and $z$ \cite{KorchFishNet}. Another approach is to replace the sum over $J$ in (\ref{AmplD6ClosedForm}) with the integral using the Watson-Sommerfeld transformation \cite{KorchFishNet}:
\begin{eqnarray}
\sum_{J} (-1)^J \mapsto \int_C\frac{dJ}{2 i \sin(\pi J)},
\end{eqnarray}
where $C$ encircles the real positive half axis in the complex $J$ plane. This allows one to safely interchange $\nu$ and $J$ integrals, and this leads us to the representation of (\ref{AmplD6ClosedForm}) in the form
\begin{eqnarray}
A^{D=6,u}(s,z,g)=\frac{1}{s}\int_C\frac{dJ}{2 i \sin(\pi J)}\int_{-\infty}^{+\infty} d\nu\frac{\mu(\nu,J)}{h(\nu,J)-g^4}\frac{\partial\Omega_{\nu,J}(z)}{\partial z}.
\end{eqnarray}
This representation turns out to be  
very useful for investigation of asymptotic behaviour of (\ref{AmplD6ClosedForm}) in the large $z$ limit - the leading large $z$ behaviour corresponds to the Regge limit and is controlled by the finite number of poles in $J$ complex plane \cite{KorchFishNet}. For finite $z$ one has to take into account an infinite series of poles, so this representation is not that helpful anymore. 

We will first focus our attention on representation (\ref{AJseriesRep}) and will try to find out what information about the weak coupling limit one can extract from it. For this aim, we found out that the following integral representation of the $\Omega_{\nu,J}(z)$ function is useful:

\begin{equation}\label{OmegaLerch}
\Omega_{\nu,J}(z)=\frac{2^J}{\pi^2}\frac{\sinh^2(\pi\nu+i\pi J/2)}{(2\pi i)^2} \int\limits_{[1,z]} \int\limits_{[1,z]} dt_1 dt_2 \frac{(t_2^2-1)^J}{2^J(t_1-z)(t_2-z)^{J+1}}\Sigma\left(\mathcal{Z},\nu,J\right),
\end{equation}
where
\begin{equation}\label{OmegaLerchSigma}
\begin{gathered}
\Sigma\left(\mathcal{Z},\nu,J\right)=-\frac{i}{2 \nu} \left( \Phi\left(\mathcal{Z},1,-\frac{J}{2} - i \nu\right)-\Phi\left(\mathcal{Z},1,-\frac{J}{2} + i \nu\right)+\right. \\
    \left.+\mathcal{Z}^{1+J}\left(\Phi\left(\mathcal{Z},1,1+\frac{J}{2} + i \nu\right) - \Phi\left(\mathcal{Z},1,1+\frac{J}{2} - i \nu\right)\right) \right),
\end{gathered}
\end{equation}
and
\begin{eqnarray}
\mathcal{Z}\equiv\frac{(t_2^2-1)(t_2-z)}{(t_1^2-1)(t_1-z)},
\end{eqnarray}
where $\Phi(z,s,\alpha)$ is the so called Lerch transcendent Zeta function \cite{LerchBook}. The $dt_1$ and $dt_2$ integration contours encircle points $1$ and $z$, while avoiding the $(-\infty,1]$ interval on the real axis in the complex $t_1$ and $t_2$ planes. Short introduction to the properties of Lerch Zeta function as well as its definition can be found in Appendix \ref{a2}. 

\subsection{Expansion around $z=-1$}\label{s41}
For the general $z$ representation (\ref{OmegaLerch}) cannot be additionally simplified, but for fixed values of $z$, namely for $z=-1$, one can obtain that
\begin{eqnarray}
\Omega_{\nu,J}(z=-1)&=&i 2^{J}~\frac{\sinh^2\left(\pi\nu+i\pi J/2\right)}{2\pi^2~\nu } 
\left(\Psi^{(0)}\left(-1-\frac{J}{2} - i\nu\right) - \Psi^{(0)}\left(\frac{J}{2} - i\nu\right) -\right. \nonumber\\
&-&\left. \Psi^{(0)}\left(-1-\frac{J}{2} + i\nu\right) + \Psi^{(0)}\left(\frac{J}{2} + i\nu\right)\right),
\end{eqnarray}
where $\Psi^{(n)}(z)$ is the PolyGamma function. It should be explicitly mentioned that in the conventions used here such a representation of $\Omega_{\nu,J}$ contains only the real part of the amplitude. This is however will be sufficient for our purposes\footnote{The question regarding appropriate analytical properties of $\Omega_{\nu,J}(z)$ as a function of $z$ is still not entirely clear for us and we would like to avoid a detailed discussion of this question here.}. In fact, all terms of the expansion around $z=-1+y$ can be represented in a similar form:
\begin{eqnarray}\label{yseries}
&&\Omega_{\nu,J}(-1+y)=\sum_{n=0} y^n~\Omega_{\nu,J}^{(n)}(-1)\nonumber\\
&&\Omega_{\nu,J}^{(n)}(-1)=\frac{i 2^{J}\sinh^2\left(\pi\nu+i\pi \frac{J}{2}\right)}{2\pi^2~\nu}P_1^{(n)}(J,\nu)
+P_2^{(n)}(J,\nu)\Omega_{\nu,J}(-1),\nonumber\\
\end{eqnarray}
where $\Omega_{\nu,J}^{(0)}(-1)\equiv\Omega_{\nu,J}(-1)$ and  $P_{1,2}^{(n)}(J,\nu)$ are some polynomials in $J$ and $\nu$. It looks rather complicated to obtain
a closed formula for these polynomials for general $n$, but for a fixed n they can be easily evaluated using Mathematica. Their explicit form for the first several values of $n$ can be found in the attached Mathematica notebook, while for $n=1$ they are 
\begin{equation}
P_{1}^{(1)}(J,\nu)=-2i(1+J)\nu,~ P_{2}^{(1)}(J,\nu)=(J(2+J)-4\nu^2)/4.
\end{equation}
Note that for $J=0$ $\Omega_{0,\nu}(z)$ is independent of $z$, which can be seen from (\ref{Ampld4Omega}), so for $J=0$ all $\Omega_{\nu,0}^{(n)}(-1)=0$ for $n>0$. Also, all additional computational details from this section can be
presented as Mathematica notebooks, if requested. 

This representation allows us to investigate behaviour of the amplitude in the vicinity of $z=-1$ in the weak coupling regime.  
Namely, we can substitute $\Omega_{\nu,J}$ in the form of (\ref{yseries}) into (\ref{AJseriesRep}) and obtain a representation of $A^{D=6,u}_4$ in the form of power series in $y=-2t/s$:
\begin{eqnarray}\label{yseries1}
&&\mbox{Re}A^{D=6,u}_4(s,-1+y,g)=\sum_{k=0}^{\infty}~\frac{c^{(k)}(g)}{s}~y^k,\nonumber\\
&&c^{(k)}(g)\sim\sum_{J=1}^{\infty}\sum_{i=1}^2\frac{2^{1-J}(J+1)((J+1)^2+4\nu^2_i)\nu_i}{(J+1)^2+4\nu_i+1}
\Omega_{J,\nu_i}^{(k)}(-1),
\end{eqnarray}
where\footnote{$\sim$ in (\ref{yseries1}) corresponds to the change in overall normalisation of $c^{(k)}(g)$ to match the conventions of Appendix \ref{a1}.} each $c^{(k)}(g)$ in turn can be expanded as power series in $g$ using the explicit expressions for $\nu$ (\ref{nu13}):
\begin{eqnarray}\label{yseries2}
\mbox{Re}A^{D=6,u}_4(s,-1+y,g)=\sum_{k=0}^{\infty}\sum_{l=0}^{\infty}~\frac{c^{(k,l)}}{s}~g^{4(l+1)}y^k,
\end{eqnarray}
where each $c^{(k,l)}$ will be given by the infinite sum over $J$. All these sums have a universal structure and can be schematically written as 
\begin{eqnarray}
c^{(k,l)}=\sum_{J=1}^{\infty}(-1)^J\sum_n \mbox{Rational function}^{(n)}(J)\times S_{n}(J),
\end{eqnarray}
where $S_{n}(J)$ is the HarmonicNumber function\footnote{Related with Polygamma via 
$\Psi^{(k)}(J)=(-1)^kk!(S_{k+1}(J-1)-\zeta_{k+1})$. For $k=0$ one has to replace the $\zeta$-function with $\gamma_E$.}. For example: 
\begin{eqnarray}\label{c01}
c^{(1,0)}=\sum_{J=1}^{\infty}(-1)^J\left(\frac{2+J+2J^2}{J^2(2+J)}+\frac{2(1+2J(2+J))}{J(1+J)(2+J)}S_{1}(J)\right).
\end{eqnarray}
To obtain such an expression, one should transform appropriate Polygamma functions in (\ref{yseries1}), according to
\begin{eqnarray}\label{PolyInverse}
\Psi^{(0)}(z)=\Psi^{(0)}(1-z)-\pi\cot(\pi z),
\end{eqnarray}
and then expand the result in powers of $g$ using the standard formulas.
Some of the $J$ sums obtained in such a way are actually finite,
as the $c^{(1,0)}$ example presented above, while the others are divergent.

However, one can make the following general observation. Using the Abel method of summation of divergent series (method of analytical continuation), it is likely that one can reconstruct all $c^{(k,l)}$. In more detail one can introduce $c^{(k),reg}(g, \tau)$: 
\begin{eqnarray}\label{yseriesReg}
c^{(k),reg}(g,\tau)=\sum_{J=1}^{\infty}\tau^J\sum_{i=1}^2\frac{2^{1-J}(J+1)((J+1)^2+4\nu^2_i)\nu_i}{(J+1)^2+4\nu_i+1}
\Omega_{J,\nu_i}^{(k)}(-1).
\end{eqnarray}
For $\tau<1$ all sums in $J$, which one can encounter expanding $c^{(k),reg}$ in $g$, will be convergent and can be expressed in terms of HPLs of the $\tau$ argument. For the first several orders of PT the analitical answers for such sums can be found in \cite{Smirnov_book_2}. The $\tau=1$ point lies on the boundary of the radius of convergence of the sums in $J$. HPLs themselves, however, can be analytically continued for $\tau \geq 1$. For example, one can take the finite $\tau \rightarrow 1$ limit in these expressions. This limit will then correctly reproduce the $c^{(k,l)}$ coefficients:
\begin{eqnarray}\label{yseriesReg1}
&&c^{(k),reg}(g,\tau)=\sum_{l=0}^{\infty}c^{(k,l),reg}(\tau)~g^{4(l+1)}\nonumber\\
&&c^{(k,l)}=\lim_{\tau \rightarrow 1^-}c^{(k,l),reg}(\tau).
\end{eqnarray}

\begin{figure}[ht]
 \begin{center}
  \epsfxsize=5cm
 \epsffile{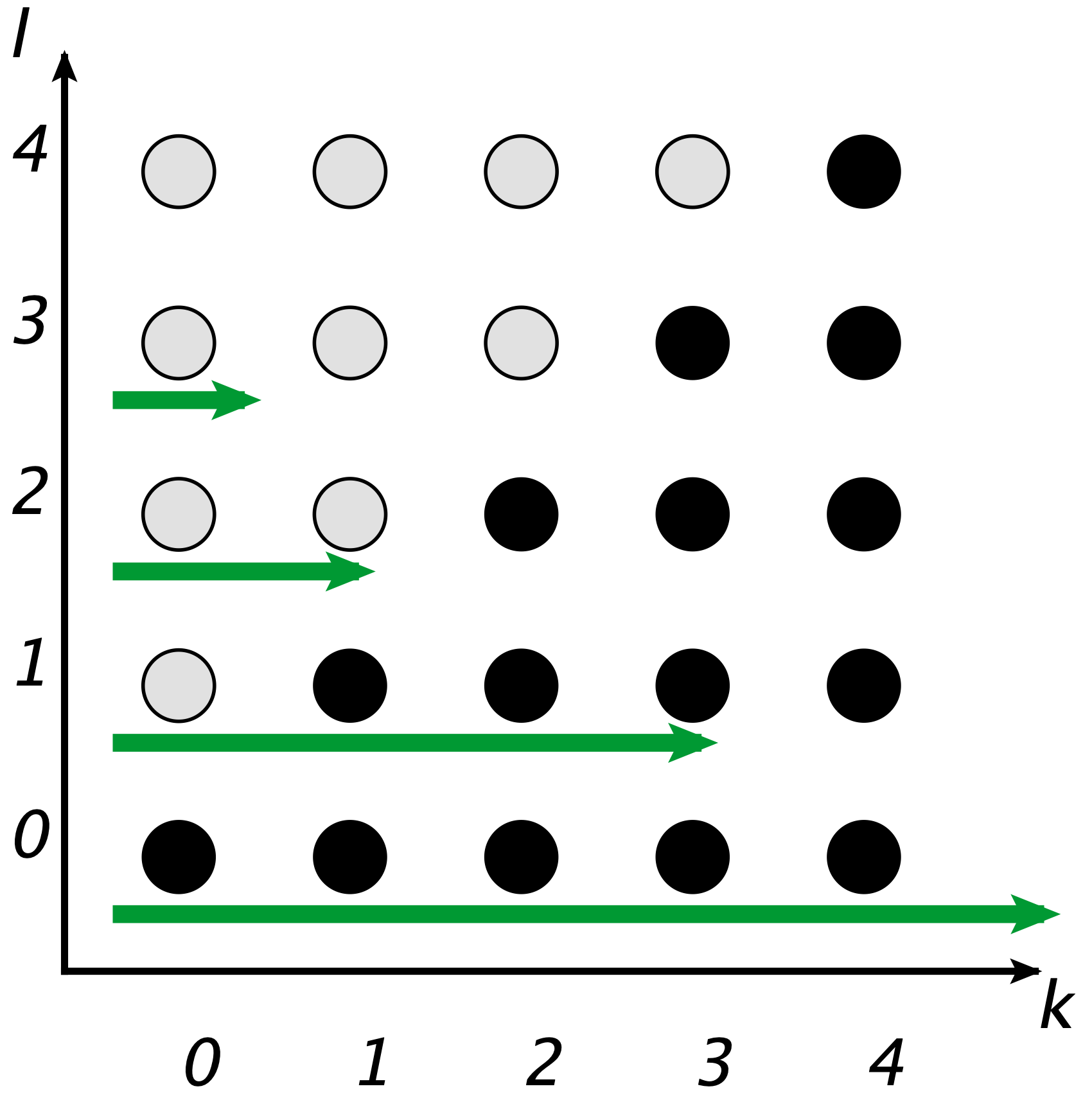}
 \end{center}\vspace{-0.2cm}
 \caption{The structure of the $c^{(k,l)}$ coefficients. Black blobs correspond to the divergent $J$ sums. Grey blobs correspond to the finite $J$ sums. Green arrows indicate the $c^{(k,l)}$ coefficients reproduced using relations (\ref{yseriesReg}) and (\ref{yseriesReg1}).}\label{fig5coef}
\end{figure}

The statement made above is strictly speaking a conjecture. We, however, have verified this conjecture by correctly reproducing the coefficients of the expansion in $\delta=y/2$ of various Box integrals listed in Appendix \ref{a1}, which include some non-trivial examples. Namely, we have correctly reproduced all tree level coefficients $c^{(k,0)}$, one loop coefficients $c^{(k,1)}$ up to $k=5$, two loop coefficients $c^{(k,2)}$ up to $k=3$ and three loop coefficient $c^{(0,3)}$. The explicit values of these coefficients can be found in Appendix \ref{a1} and the explicit form of the corresponding series are given in the attached Mathematica notebook.
For example, the whole tree level series of the $c^{(k,0)}$ coefficients can be reconstructed and are given by geometric progression
\begin{eqnarray}
c^{(k,0),reg}&=&\sum_{J=1}^{\infty}(-\tau)^J\left(\frac{(J-k)\ldots(J-1)(J+k+1)\ldots(J+2)(1+2J)}{(k+1)!k!}\right),\nonumber\\
c^{(k,0)}&=&c^{(k,0),reg}(\tau=1)=\frac{(-1)^k}{2^k},
\end{eqnarray}
which in turn correctly reproduce expansion of the $B^{(0)}(s,u)=1/u$ tree level diagram in powers of $y$. Indeed $1/u=1/s \times 2/(2+y)$. This may look somewhat incidental. But the correct reproduction of $c^{(4,1),Reg.}(1)=c^{(4,1)}= 11/24$, which is the coefficient of $y^4$ in the one loop Box integral expansion, and which is given by the following (divergent for $\tau=1$) series
\begin{eqnarray}
c^{(4,1),reg}&=&\sum_{J=1}^{\infty}(-\tau)^J\left(\frac{(-2+J)(-1+J)(3+J)(4+J)(-3-8J+4J^3+J^4)}{36(1+J)(2+J)J}S_{1}(J)+\right.\nonumber\\
&+&\left.\frac{(-2+J)(-1+J)(-432-72J+87J^2+571J^3+516J^4+169J^5+19J^6)}{432J^2(2+J)}\right),\nonumber\\
c^{(4,1)}&=&c^{(4,1),reg}(\tau=1)=\frac{11}{24}
\end{eqnarray}
looks non-trivial. We have also verified that these results are stable under the change of the summation method. Namely, we have reproduced the tree level results in $\zeta$ regularisation. As another example, we will list here the sum for $c^{(2,2)}$ that gives the coefficient of $y^1$ in the two loop Box expansion:
\begin{eqnarray}
&&c^{(2,2)}=\sum_{J=1}^{\infty}(-1)^J\left(\frac{\left(-J^4-4 J^3+J^2+10 J+4\right)}{J^2 (J+1)^2 (J+2)^2}S_2(J)+\right.\nonumber\\
&+&\frac{2 \left(2 J^8+16 J^7+43 J^6+34 J^5-31 J^4-52 J^3-2 J^2+20 J+8\right)}{J^3(J+1)^3(J+2)^3}S_1(J)-\nonumber\\
&-&\frac{2 \left(12+\pi ^2\right) J^{10}+\left(210+19 \pi ^2\right) J^9+\left(726+71 \pi
^2\right) J^8}{6 J^4 (J+1)^4 (J+2)^3}-\nonumber\\
&-&\frac{\left(1221+125 \pi ^2\right) J^7+\left(861+79 \pi ^2\right) J^6-8 \left(138+17
\pi ^2\right) J^4}{6 J^4 (J+1)^4 (J+2)^3}-\nonumber\\
&-&\left.\frac{-2 \left(171+32 \pi ^2\right) J^5-4 \left(153+20 \pi ^2\right) J^3-8 \left(2
\pi ^2-27\right) J^2+336 J+96}{6 J^4 (J+1)^4 (J+2)^3}\right)=\nonumber\\
&=&\frac{1}{2}+\frac{\pi^2}{2}-\frac{11\pi^4}{180}.
\end{eqnarray}
In addition, we want to mention that the whole sequence of coefficients $c^{(0,l)},~ l\geq1$ is given by the finite sums. 
See fig. \ref{fig5coef}.

We have also tried to probe different kinematic points, namely $z=1$, using a similar method\footnote{This can be done because the $\Omega_{\nu,J}(z)$ functions at the points $z=-1$ and $z=1$ are related as $\Omega_{\nu,J}(-1)=(-1)^J\Omega_{\nu,J}(1)$.} of summation, but found no agreement with the explicit results of Box integrals and results of the Abel summation method. For $z=1$ the Abel summation is no longer applicable because of singular $\tau=1$ limit, which is probably related to the fact that around $z=-1$ all series in $J$ are sign alternating, while for  $z=1$ they are not. 

However, we have observed that the series of coefficients $a^{(0,l)}$, which are given by the $\mathcal{B}^{(l)}(x)$ functions evaluated at $x=t/s=0$, starting from two loops are given by the finite sums. This allows us to reproduce some results of \cite{PanzerHyperInt} and obtain a generating function for these coefficients (which is a special case of the Box generating function):
\begin{equation}\label{GeneratingFunctionD6Box}
\sum_{l=2}^{\infty}\mathcal{B}^{(l)}(0)~g^{4l}=\sum_{l=2}^{\infty}a^{(0,l)}~g^{4l}\sim\frac{\partial^3}{\partial (g^{4})^3} \int_{-\infty}^{+\infty} d\nu\sum_{J \geq 0}^{\infty}(-1)^J\frac{\mu(\nu,J)}{h(\nu,J)-g^4}\frac{\partial\Omega_{\nu,J}(-1)}{\partial z}.
\end{equation}
Here the derivative with respect to the coupling constant stands for the purpose of cancelling the first couple of divergent terms in weak coupling expansion. 
The first pair of finite $a^{(0,l)}$ coefficients, which can be extracted from (\ref{GeneratingFunctionD6Box}), is:
\begin{eqnarray}\label{GeneratingFunctionD6Box1}
a^{(0,2)}&=&2\zeta_2,\nonumber\\
a^{(0,3)}&=&4\zeta_3^2+\frac{124}{35}\zeta_2^3-8\zeta_3-6\zeta_2.
\end{eqnarray}
One can see that $a^{(0,l)}$ are given by the same $J$ sum expressions as their $c^{(0,l)}$ counterparts but without the overall $(-1)^J$ coefficient. 
These coefficients at high orders of PT ($l \leq6$) were first computed in \cite{PanzerHyperInt} in an analytical form using the approach of direct evaluation of the Feynman parameter integrals and HyperInt Maple library \cite{PanzerHyperInt}. It is interesting whether it is possible to obtain the closed expression (\ref{GeneratingFunctionD6Box}) for these coefficients using the methods of \cite{PanzerHyperInt}. 

The results of the previous discussion combined with known functional space which one works with at $l$-loops, can be actually used to bootstrap perturbative results (similar to \cite{ExactCorrelationFunctionsFishnet}) and to reconstruct the $B^{(l)}(s,u)=\mathcal{B}^{(l)}(x)/s$, $x=u/s$ Box integral and hence the $A^{D=6}$ amplitude at $l$-loops. One can write the ansatz of the form
\begin{equation}
\mathcal{B}^{(l)}(-1+y)=\frac{\sum_{\vec{a}} C^{(1)}_{\vec{a}}H_{a_1,\ldots a_{w_2l}}(-1+y)}{y}+
\frac{\sum_{\vec{a}} C^{(2)}_{\vec{a}}H_{a_1,\ldots a_{w_2l}}(-1+y)}{-1+y},
\end{equation}
where $C^{(1,2)}_{\vec{a}}$ are arbitrary coefficients. Then one can compare expansion of this ansatz in powers of $y$ with (\ref{yseries2}). With enough powers of $y$ expansion
considered, a system of linear equations for all $C^{(1,2)}_{\vec{a}}$ can be obtained.
We have reproduced one and two loop results using this approach. 
All results and approaches mentioned above are not specific to the $D=6$ case but can be applied to the $D=4$, $\omega=1$ theory. Additionally, different kinematic limits can also be used to generate additional constraints on $C^{(1,2)}_{\vec{a}}$. We will discuss one of these limits below.

\subsection{Expansion around $z=\infty$}\label{s42}
It has been well known for a long time that high energy behaviour of amplitudes is usually simplified and is controlled by the Regge theory \cite{GribovBook,DiffractionBook}. So it is useful to consider a similar limit in our case. We will consider the
$u \gg s$ limit which translates to $z \gg 1$ in our notation. As was mentioned before, such a limit is controlled by the Regge theory. Note that kinematic points, considered in the previous section, can be associated with high energy behaviour as well, because $z=-1$ and $z=1$ map to the $s \gg t$ limits $s \gg u$ respectively. 

In the large $z$ limit, each order of PT of the amplitude $A^{D=6}$ is dominated by powers of $\log(x) \equiv L$, $z\sim x=u/s$ \cite{GribovBook,KorchFishNet}: 
\begin{eqnarray}
s~A^{D=6}&\sim& s~A^{D=6,u}=\sum_{l}(g^4)^{l+1} \mathcal{B}^{(l)}(x),\nonumber\\
x~\mathcal{B}^{(l)}(x)&=&a_{(l)}^{LLA}L^{2l}+a_{(l)}^{NLA}L^{2l-1}+a_{(l)}^{NNLA}L^{2l-2}+\ldots,
\end{eqnarray}
where $a_{(l)}^{N^kLA}$ are some numerical coefficients.
In \cite{KorchFishNet} a technique was developed on how to extract the coefficients $a^{N^kLA}_{(l)}$. This technique was based on the asymptotic representation of the $\Omega_{\nu,J}(z)$ function for large $z$ in the form 
\begin{equation}\label{Ampld4_1}
\Omega_{\nu,J}(z)=(-1)^J\frac{\sinh(2\pi\nu)}{2\pi\nu}
\frac{\Gamma(J-2i\nu+1)\Gamma(J+2i\nu+1)}{\Gamma(J/2-i\nu+1)^2\Gamma(J/2+i\nu+1)^2}z^J+O(z^{J-2}).
\end{equation}
This allows one to write an approximate expression for the amplitude which takes into account all log enhanced contributions:
\begin{equation}\label{1}
s~A^{D=6,u}(s,z,g)\sim\int^{g^2}_{-g^2}d\nu\left( J_{+}F(\nu,J_{+})z^{J_{+}-1}-J_{-}F(\nu,J_{-})z^{J_{-}-1}\right)+\ldots,
\end{equation}
with
\begin{eqnarray}
J_{\pm}&=&-1+\sqrt{1-4\nu^2\pm 4\sqrt{g^4-\nu^2}},\nonumber\\
F(\nu,J)&=&\frac{\nu \sinh^2(2\pi\nu)\Gamma(J-2i\nu+2)\Gamma(J+2i\nu+2)}{\sin(\pi J)(J(J+2)+4\nu^2)\Gamma(J/2-i\nu+1)^2\Gamma(J/2+i\nu+1)^2}.
\end{eqnarray}

This expression allows one to obtain the coefficients $a_{(l)}^{N^kLA}$ up to arbitrary orders of perturbation theory $l$. In our case, following similar steps as in \cite{KorchFishNet}, we can obtain for the first several coefficients:
\begin{equation}\label{aLLA}
a_{(l)}^{LLA}=\frac{1}{l!(l+1)!},
\end{equation}
\begin{equation}\label{aNLA}
a_{(l)}^{NLA}=\frac{2l(l-1)}{l!(l+1)!},
\end{equation}
\begin{equation}\label{aNNLA}
a_{(l)}^{NNLA}=\frac{2l(l-1)(l+2)+\pi^2(l+1)}{l!(l+2)!},
\end{equation}
and
\begin{equation}\label{aNNNLA}
a_{(l)}^{NNNLA}=\frac{2l \left(l (l+1) \left(2 l^2+2 l+3 \pi ^2-13\right)+6 (l-4) \zeta_3+18\right)}{3l!(l+2)!}.
\end{equation}
All are in perfect agreement with one, two and three loop computations. See Appendix \ref{a1}. We want to stress that these coefficients are evaluated in an arbitrary number of loops $l$.

So let us summarise what we have learned so far. We have seen that representation (\ref{AmplD6ClosedForm}) is consistent with the results of the standard PT based on the Feynman diagram expansion and even allows one to extract some non-trivial information, like the coefficient (\ref{aNNNLA}), about all orders of PT expansion of the amplitude. In fact, (\ref{AmplD6ClosedForm}) allows one to reconstruct a full answer in a fixed number of loops, though not in a straightforward manner\footnote{It is doubtful that the bootstrap approach described here will be efficient at high loop orders due to the factorial growth of dimensionality of functional space of HPLs and that this approach will be more efficient than the HyperInt library \cite{PanzerHyperInt}, which however also operates with expressions which grow very fast with the number of loops. Most likely, both approaches will be in fact limited by the ability of current mainstream software (Mathematica or Maple) to efficiently manipulate large symbolical expressions.}. 

\section{$A_4$ amplitude at strong coupling}\label{s5}
Representations of the amplitudes such as (\ref{Ampld4}) and (\ref{AmplD6ClosedForm}) are valid for arbitrary value of the coupling constant. This means that they allow one to investigate properties of the amplitude in the strong coupling limit, which is obviously unavailable from the weak coupling Feynman diagram expansion. In this section, we are going to investigate the behaviour of the amplitude in this limit. Since the $D=4$ and $D=6$ amplitudes are related via simple differentiation, for simplicity we will consider mainly the $D=4$ case. As a starting point, it is useful to consider the amplitude after the Watson-Sommerfeld transformation. We want to define the strong coupling limit 
such that\footnote{This is necessary to make $\nu$ integrals in formulas like (\ref{Ampld4}) and (\ref{AmplD6ClosedForm}) well defined \cite{ExactCorrelationFunctionsFishnet}. One should add a small imaginary part to $g$ to avoid poles on the real axis in $\nu$ integral. Alternatively, one can try to deform the $\nu$ integration contour. This turns out to be irrelevant in the weak coupling regime, but is important in the strong coupling case.} $g\in \mathbb{C}$, $|g|\gg1$, $J\sim |g|$ i.e. we will rescale $J\mapsto g~J$ and also consider $\mbox{Im}\:g>0$. The requirement $J\mapsto g~J$ was motivated by 
the results of \cite{ExactCorrelationFunctionsFishnet} for correlation functions where the main contribution to the correlation functions in strong coupling was coming from large $J$ in the sum (\ref{4PointCorrExact}).
Overall this allows us to write the amplitude in the following way:
\begin{eqnarray}
A^{D=4,u}(z,g)=\frac{1}{2 i}\int_C\frac{d (g J)}{\sin(\pi gJ)}\int_{-\infty}^{+\infty} d\nu\frac{\mu(\nu,gJ)}{h(\nu,gJ)-g^4}\Omega_{\nu,gJ}(z),
\end{eqnarray}
where the $\Omega_{\nu,J}(z)$ function is chosen in representation (\ref{OmegaLerch}). The $\Omega_{\nu,J}(z)$ function contains a common $\sinh$ factor which in the $|g|\gg1$ limit generates 
exponentially enhanced terms of the form $\exp(+|g|f(J))\ldots$. This hints that the steepest descent method can be used for evaluation of $J$ or $\nu$ integrals. 
So our strategy for investigation of the strong coupling limit consists of an attempt to simplify the integrand of (\ref{OmegaLerch}) in the $|g|\gg1$ limit and the subsequent evaluation of the $\nu$ integral by residues and $J$ integral by the steepest descent method. This approach is similar to the results of \cite{ExactCorrelationFunctionsFishnet} for the correlation functions.

With the constraints mentioned above one can remove the $\nu$ integral by taking residues in the complex 
$\nu$ plane. The poles are defined by the equation:
\begin{equation}
h(\nu,g J) = \left(\nu^2 + \frac{(g J)^2}{4}\right) \left(\nu^2 + \frac{(g J + 2)^2}{4}\right)=g^4. 
\end{equation}
This equation will have four roots (identical to the weak coupling case) which are given by
\begin{eqnarray}
\nu_{1,2}&=&\pm \frac{1}{2} \sqrt{-2 - 2 g J - g^2 J^2 + 2 \sqrt{1 + 4 g^4 + 2 g J + g^2 J^2}},\nonumber\\
\nu_{3,4}&=&\pm \frac{1}{2} \sqrt{-2 - 2 g J - g^2 J^2 - 2 \sqrt{1 + 4 g^4 + 2 g J + g^2 J^2}}.
\end{eqnarray}
We will close the contour in the lower half of the complex plane and with our conventions this will
encircle the $\nu_2$ and $\nu_4$ roots. 
For convenience, hereafter we will replace 
\begin{equation}
g\mapsto \frac{1}{g},~g \rightarrow 0
\end{equation}
in all expressions. With these conventions the leading behaviour of the $\nu$ roots as a function of $J$
are given by: 
\begin{equation}
\begin{gathered}
    \nu_1(J) = \frac{\sqrt{4 - J^2}}{2 g} - \frac{J}{2 \sqrt{4 - J^2}} + O(g), \\
    \nu_2(J) = -\nu_1(J), \\
    \nu_3(J) =  \frac{i \sqrt{4 + J^2}}{2 g}+  \frac{i J}{2 \sqrt{4 + J^2}} + O(g),\\
    \nu_4(J) = - \nu_3(J).
\end{gathered}
\end{equation}
After residue evaluation the amplitude can be written as:
\begin{eqnarray}\label{AmplitudeStrongJint}
    A_4^{D=4,u}(z,1/g) =  8\pi \sum_{\nu=\nu_2, \nu_4} \frac{1}{2 i}\int_C\frac{d (J/g)}{\sin(\pi J/g)}\frac{2(J/g+1)(4\nu^2+(J/g+1)^2)\nu}{(4 \nu^2+(J/g+1)^2+1)}\frac{\Omega_{\nu,J/g}(z)}{2^{J/g}}.
\end{eqnarray}

One can note that in the limit $1/g \gg 1$ the $\Omega_{\nu,J/g}(z)$ function or more precisely the 
integrand of the $\Omega_{\nu,J/g}(z)$ function in the integral representation (\ref{OmegaLerch}) can be simplified significantly. Namely,
one can transform two Lerch Zeta functions in $\Phi(z,1,a)$, according to \cite{LerchInv}, which may be seen as a generalisation of relation (\ref{PolyInverse}) used in the weak coupling regime:
\begin{eqnarray}
\Phi(z, 1, a)= z^{-1}\Phi(z^{-1}, 1,1-a)+\pi z^{-a}(\cot(\pi ~a)- i \: \text{sgn}(\varphi)),
\end{eqnarray}
where $\varphi=\arg(-\ln{z})$, so that in all four $\Phi(z,1,a)$ functions $\mbox{Re}~a>0$, and then use the following asymptotic expansion for each of them
\begin{eqnarray}
    \Phi(z, 1, a) = \frac{1}{1-z}\frac{1}{a} + O(a^{-2}).
\end{eqnarray}
Then the other parts of the integrand in (\ref{AmplitudeStrongJint}) can be expanded in powers of $g$ and the integrals in $t_1$ and $t_2$ in the representation of $\Omega_{\nu,J/g}(z)$ can
be evaluated by residues. After that, the result is given by the following asymptotic expression for the amplitude:
\begin{equation}\label{AmplitudeStrongJintFin}
\begin{gathered}
    A^{D=4,u}_4(z,1/g) = \frac{1}{2 \pi i (\sqrt{z^2-1})}  \int\limits_{C}\frac{d (J/g)}{\sin{(\pi J/g)}} J \sqrt{4 - J^2} \exp{\left(\frac{\pi \sqrt{4-J^2}}{g}+(1+\frac{J}{g})\mathbf{L}\right)}+\ldots,
\end{gathered}    
\end{equation}
where $\ldots$ corresponds to all terms suppressed by powers of $g$ or $\exp(-1/g)$ and 
\begin{equation}
\mathbf{L} \equiv \log{(z+\sqrt{z^2-1})}.
\end{equation}
All intermediate steps,
including the explicit expressions which give rise to suppressed terms, can be found in Appendix \ref{a3}. It
should also be mentioned that the expression above is valid only if the following conditions for $ z \ll \text{e}^{\sqrt{2}\pi/\sqrt{g}}$ and $z>1$ are satisfied.

The $J$ integral in (\ref{AmplitudeStrongJintFin}) can then be evaluated by the steepest descent method around the point
\begin{equation}
    J_0 =  \frac{2 \mathbf{L}}{\sqrt{\mathbf{L}^2+\pi^2}},
\end{equation}
which results in the following strong coupling limit for the amplitude:
\begin{equation}\label{AmplStrongCouplingFinal}
    A^{D=4,u}_4(z,1/g) =g^{-1/2} \frac{4\pi\: \pi^{1/2} \: \mathbf{L} \exp{\left(\frac{2}{g}\sqrt{\pi^2+\mathbf{L}^2}\right)}}{i \sqrt{z^2-1} \: (\pi^2+\mathbf{L}^2)^{7/4} \sin{\left( \frac{2\pi \mathbf{L}}{g \sqrt{\pi^2+\mathbf{L}^2}}\right)}}+\ldots,
\end{equation}
while the $D=6$ counterpart can be obtained from this expression as $s A^{D=6,u}_4=\partial/\partial z ~A^{D=4,u}_4$. 

Following similar steps one can consider the $z \gg \text{e}^{\sqrt{2}\pi/\sqrt{g}} \gg 1$ situation which corresponds to the Regge limit. The result for the amplitude in this limit will be given by:
\begin{equation}\label{AmplStrongCouplingFinalRegge}
    A^{D=4,u}_4(z,1/g)  \sim  \frac{z^{2/g-1}}{ \log^{3/2}(z)}+\ldots,
\end{equation}
which is consistent with the previous results \cite{KorchFishNet}. Both expressions (\ref{AmplStrongCouplingFinal}) and (\ref{AmplStrongCouplingFinalRegge}) for $z\gg 1$ have identical Regge behaviour $\sim z^{2/g-1}$ with the Regge trajectory $J_R=2/g-1+\ldots$ but different pre-exponential factors. It  is also worth mentioning that using (\ref{aLLA}), one can \cite{KorchFishNet,BorkD6} reconstruct the leading weak coupling expansion for the
Regge trajectory $J_R=2g^2+\ldots$ and the exact results for the Regge trajectory can be obtained from (\ref{Ampld4_1})
\begin{eqnarray}
J_{R}^{D=4}(g)&=&\sqrt{1+4g^2}-1,\\
J_{R}^{D=6}(g)&=&\sqrt{1+4g^2}-2,
\end{eqnarray}
which interpolates nicely between the weak and strong coupling results.

Expression (\ref{AmplStrongCouplingFinal}) is not valid for $z=\pm1$ points. This point however can be investigated separately using a similar
integral representation for the $\Omega_{\nu,J}(z)$ function. The resulting answer will be
\begin{equation}\label{AmplStrongCouplingFinalz1}
A^{D=4,u}_4\left(z=\pm1,1/g\right) = \sqrt{g} \frac{ 8\pi}{2\pi^3 i}  \exp{\left(2 \pi/g\right)} +\ldots.
\end{equation}

It would be interesting to reproduce the results obtained above by some dual gravitational model similarly to the famous quasi-classical  $AdS_5$ minimal volume computations \cite{ampWLduality1}. 

\section{Relations to other theories}\label{s6}
It is interesting to make several observations regarding the relations of other different $D=6$ models with the results obtained here. First of all, it looks like the $D=6$, $\omega=1$ fishnet model considered here gives some (uncontrolled) approximation to the planar $\mathcal{N}=(1,1)$ SYM theory, at least at the level of four-point amplitudes. Also, it is likely that Regge behaviour of both theories may be related to one another.
Indeed, the four-point single-trace partial colour ordered amplitude $A_4^{SYM}$ in $\mathcal{N}=(1,1)$ SYM at first several orders of PT is given only by the $D=6$ Box functions \cite{Dennen:2010dh,BorkD6}:
\begin{equation}
\frac{A_4^{SYM}}{A_4^{SYM,tree}}=1+g_{YM}N_c\frac{st}{2}B^{(1)}(s,t)+(g_{YM}N_c)^2\left(\frac{s^2t}{4}B^{(2)}(s,t)+\frac{st^2}{4}B^{(2)}(t,s)\right)+\ldots.
\end{equation} 
Higher orders of PT will also always contain a pair of $B^{(l)}(s,t)$ and $B^{(l)}(t,s)$ diagrams. The same pair of Box diagrams also will always appear at the double-trace level in each order of PT. High energy limit of $A_4^{SYM}$ at a single-trace level will likely be controlled by the series of $B^{(l)}(s,t)$ or $B^{(l)}(t,s)$ diagrams \cite{Bork:2015zaa,Bork:2014nma,BorkD6}. The behaviour of such a series of Box diagrams is governed by the fishnet model studied here. It is interesting that the non-renormalizable and non-conformal theory such as $\mathcal{N}=(1,1)$ SYM still contains, in some sense, a CFT subsector within itself.

It is also interesting to note the relation of the results obtained here with \cite{Lip}. In \cite{Lip} two questions were raised. The first one is which scalar integrals can contribute to the colour ordered amplitude with dual conformal symmetry in the hypothetical $D=6$ gauge theory. The second question was how to compute the minimal surface in $AdS_7$ with a boundary that forms a polygonal contour with four cusps and is attached to $\partial AdS_7$.
It turned out that the answers to both questions are related to each other. The only one 
loop scalar integral which can contribute to the hypothetical amplitude must include the $1/p^4$ propagators and is given by fig.\ref{fig1} B) (its momentum space version with $p_i^2=0$ on shell condition for external legs to be exact). The minimal area, in its turn,
was given, roughly speaking, by the exponent of this one loop integral \cite{Lip}. This all mimics the $D=4$, $\mathcal{N}=4$ SYM case. In $\mathcal{N}=4$ SYM we have the BDS ansatz answer for the four-point amplitude \cite{BDS1} which is also roughly speaking given by the exponent of one loop result and in the  strong coupling regime is consistent with the $AdS_5$ minimal surface computation \cite{ampWLduality1}. This allows the authors of \cite{Lip} to speculate that their $D=6$ results are related to the $\mathcal{N}=(2,0)$ SYM theory (see, for example, \cite{Lambert20Reviev,Lambert:2012qy,Douglas:2010iu,Lip} and references therein). In \cite{BKIT}, the question of systematic construction of the $D=6$ scalar integrals with $1/p^4$ propagators with iterative structure, which is consistent with $AdS_7$ minimal surface area computations \cite{Lip}, was investigated. The algorithm for constructing such $D=6$ integrals was suggested \cite{BKIT} and the $D=6$ hypothetical amplitude candidate $M_4^{D=6,(l)}$ with iterative structure turned out to be given at the $l$-loop level by the derivatives of the $D=4$ $\mathcal{N}=4$ SYM amplitude $M_4^{D=4,(l)}$:
\begin{equation}
M_4^{D=6,(l)}=\left(t\frac{\partial}{\partial t}+s\frac{\partial}{\partial s}-(l+1)\right)M_4^{D=4,(l)}.
\end{equation}
This is very similar to the relation  (\ref{AmplD6ClosedForm}) between the $D=4$ and $D=6$ fishnet amplitudes obtained here.
This in turn allows us to make speculations about the possibility that the parent theory to the 
$D=6$, $\omega=1$ fishnet considered here is the $D=6$, $\mathcal{N}=(2,0)$ SYM theory similar to the $D=4$, $\omega=1$ fishnet and $\mathcal{N}=4$ SYM counterparts.

In conclusion, the results of investigation of the $D=6$ $\mathcal{N}=(1,0)$ SYM higher derivative theories should be mentioned \cite{Buchbinder:2020ovf,Buchbinder:2020tnc,Buchbinder:2019goz,Merzlikin:2018tbv,Buchbinder:2018bhs}.
It would be interesting to investigate the possibility of constructing the $D=6$ $\mathcal{N}=(1,0)$ SYM CFT model and its possible relations with the fishnet CFT studied here. 

\section{Conclusions}\label{s7}
In this article the four-point scattering amplitudes in the $D=4$ and $D=6$ fishnet CFT models were investigated.
It was pointed out that the $D=6$, $\omega=1$ fishnet model is related to its $D=4$ counterpart at the level of four-point amplitudes via a derivative with respect to one of the Mandelstam invariants. This allows us to investigate both models simultaneously. Also, as a by product of such investigation, the generating function for the $D=6$ on-shell Box ladder diagrams was obtained. In the weak coupling limit we have successfully reproduced various terms of the expansion of the $D=6$, $\omega=1$ amplitude around the $z=-1$ ($t=0$) kinematic point. This also allows us to suggest the bootstrap method for computation of the four-point $l$-loop amplitude or equivalently the Box ladder diagrams with $l-1$ rungs. In addition, some all-loop results on logarithmic asymptotics of Box ladder diagrams were obtained. Also, the new integral representation of the $\Omega_{\nu,J}$ function was suggested which allows us to obtain strong coupling leading asymptotic of four-point scattering amplitudes in the $D=4$ and $D=6$, $\omega=1$ models. This combined with PT results for both models gives us a full qualitative description of the four-point scattering amplitude in such models for arbitrary kinematics and all values of the coupling constant.  

Also, possible relations of the $D=6$, $\omega=1$ fishnet with other $D=6$ theories were discussed. The results obtained here may be related in various ways to other $D=6$ gauge
theories with (extended) supersymmetry similar to the $D=4$, $\omega=1$ fishnet model and $\mathcal{N}=4$ SYM relations.

There are several directions of possible further developments which we find interesting. It would be 
interesting to investigate systematically strong coupling limit corrections to the leading asymptotic expression for the $D=4$ amplitude found here. This may be relevant for the construction and investigation of the dual string description of fishnet models in $D=4$ and, probably, in higher dimensional cases. In addition, it would be interesting to systematically investigate the spectrum of anomalous dimensions of
local operators in such a theory. This may give us hints about the parent theory for the $D=6$
fishnet considered here.
It would also be interesting to generalise the results obtained here for four-point amplitudes to the higher numbers of external particles. The 6-point amplitude is of particular interest because such results should be related to the $\mathcal{N}=4$ SYM NMHV$_6$ amplitude, which is an important object for investigation by various integrability-based approaches \cite{Papathanasiou:2013uoa,Drummond:2015jea,Cordova:2016woh,Bork:2019aud,Basso:2020xts}. 

\section*{Acknowledgements}
L.V. is grateful to A.I.Onishchenko and D.I.Kazakov for useful discussions.
E.S. thanks A.M. Fedotov for fruitful discussions.
The work of L.V. Bork was supported by the Foundation for the
Advancement of Theoretical Physics and Mathematics "BASIS". The R.M Iakhibbaev's work is supported by a RSF grant no. 21-12-00129.
\newpage
\appendix
\section{$D=6$ box integrals}\label{a1}
\begin{figure}[ht]
 \begin{center}
  \epsfxsize=17cm
 \epsffile{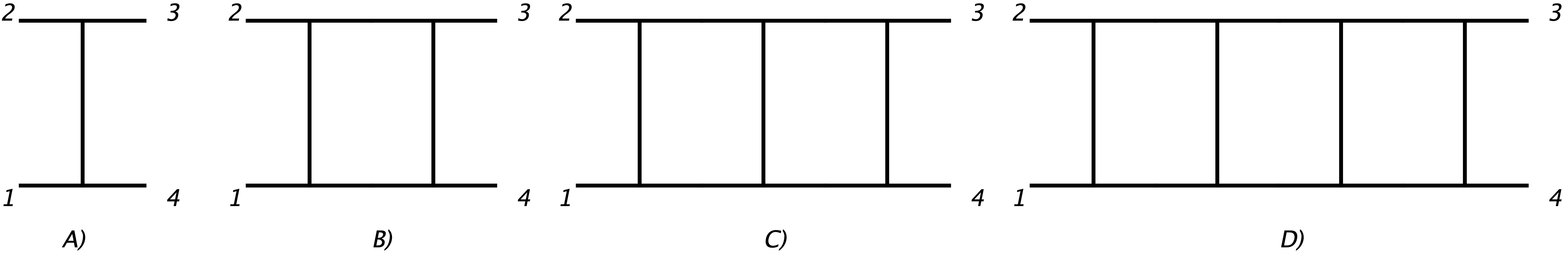}
 \end{center}\vspace{-0.2cm}
 \caption{$D=6$ Box integrals.}\label{fig11a}
 \end{figure}
With appropriate normalization the result of the $l$-loop $D=6$ Box function $B^{(l)}(s,t)$ for any $l$ can be written as
\begin{eqnarray}
B^{(l)}(s,t)=\frac{1}{s}\left(\frac{a^{(l)}(x)}{1+x}+\frac{b^{(l)}(x)}{x}\right),
\end{eqnarray}
where $a^{(l)}(x)$ and $b^{(l)}(x)$ are dimensionless functions of single argument $x=t/s$.
It is convenient to consider the rescaled Box function $\mathcal{B}^{(l)}(x)$:
\begin{eqnarray}
B^{(l)}(s,t)=\frac{\mathcal{B}^{(l)}(x)}{s}.
\end{eqnarray}
The $u$-channel
Box function will be given by the same expression but with $x=u/s$.
Hereafter we will list only the $a^{(l)}(x)$ and $b^{(l)}(x)$ functions. The full result can be reconstructed by the given above formulas.

The tree level $l=0$ (see fig.\ref{fig11a} A) result can be defined simply as
\begin{eqnarray}
a^{(0)}(x)&=&0,\nonumber\\
b^{(0)}(x)&=&1.
\end{eqnarray}

The one loop Box (see fig.\ref{fig11a} B) can be evaluated by multiple methods and the result is given by
\begin{eqnarray}
a^{(1)}(x)&=&\frac{\log^2(x)}{2}+\frac{\pi^2}{2},\nonumber\\
b^{(1)}(x)&=&0.
\end{eqnarray}

At any loop order $l$ the $D=6$ Box function ($a^{(l)}(x)$ and $b^{(l)}(x)$) can be expressed in terms of the so called Harmonic Polylogarithms \cite{HPL_init}. For completeness let us briefly discuss their definition and some of their main properties.
Let us consider $x\in (0,1)$ and the set of numbers $a_i \in \{-1,0,1\}$. Then by definition we have
\begin{equation}
    H_0(x) = \log(x); \quad H_{\pm 1} = \mp \log (1\mp x ),
\end{equation}
and
\begin{equation}
    \frac{d}{dx} H_a (x) = f_a(x),
\end{equation}
where
\begin{equation}
    f_0(x) = \frac{1}{x}; \quad f_{\pm 1} = \frac{1}{1\mp x}.
\end{equation}
Then the $H_{a_1,...,a_n}(x)$ Harmonic Polylogarithm (HPL) function can be defined iteratively for any given set of $a_i$'s:
\begin{equation}
    H_{a_1,...,a_n}(x) = \int_0^x dx' f_{a_1} (x') H_{a_2,...,a_n} (x'), 
\end{equation}
with
\begin{equation}
    H_{0,...,0} (x) = \frac{1}{n!} \log^n(x).
\end{equation}
The derivative of the HPL is defined as
\begin{equation}
    \frac{d}{dx} H_{a_1,...,a_n} (x) =  f_{a_1} (x) H_{a_2,...,a_n} (x).
\end{equation}
The HPLs can be analytically continued from $x\in (0,1)$ to the full real axis. 
The number of $a_i$ indices is called the weight of HPL. For example, if the weight is equal to $n$, then we have $3^n$ different but functionally dependent HPL functions.
If we have a string of $(k-1)$ zeroes and then $\pm 1$, such a combination of indices will be labelled 
\begin{equation}
    0,...,0,\pm1 \rightarrow \pm k.
\end{equation}
For example:
\begin{equation}
    H_{0,0,0,1,0,0,-1,0,1} = H_{4,-3,2}.
\end{equation}
The product of two HPL functions can be lineralized according to the relation
\begin{equation}
    H_{\vec{a}} (x) H_{\vec{b}}(x) = \sum_{\vec{c}=\text{Perm}(\vec{a},\vec{b})} H_{\vec{c}}(x),
\end{equation}
where the sum runs through all possible permutations from $\{a_i,b_j\}$, which leaves the order of arguments
unchanged.
For example,
\begin{equation}
    H_{a}(x) H_{b_1,b_2}(x) = H_{a,b_1,b_2}(x) + H_{b_1,a,b_2}(x) + H_{b_1,b_2,a}(x).
\end{equation}
It is easy to see that for a pair of HPLs with weights $n$ and $m$ the product will have weight $(n+m)!/(n! m!)$. For saving space, hereafter we will also drop the explicit indication of the argument of HPL: $H_{a_1,\ldots,a_n} \equiv H_{a_1,\ldots,a_n}(x)$.

In these notations the one-loop result can be rewritten as
\begin{eqnarray}
a^{(1)}(x)&=&H_{0,0}+\frac{\pi^2}{2},\nonumber\\
b^{(1)}(x)&=&0.
\end{eqnarray}

The two-loop box (see fig.\ref{fig11a} C)  can be evaluated by the differential equation technique \cite{D6Box2loops}, Mellin-Barnes representation \cite{BorkD6} or by directly integrating the Feynman parameters \cite{PanzerHyperInt}. The result is given by:
\begin{eqnarray}
a^{(2)}(x)&=&-\pi^2H_{-1}+\frac{\pi^2}{3}H_{0}-2H_{-1,0,0}-2\zeta_3,\nonumber\\
b^{(2)}(x)&=&\pi^2H_{-1,-1}-\frac{\pi^2}{3}H_{-1,0}+2H_{-1,-1,0,0}+2\zeta_3H_{-1}.
\end{eqnarray}

The three-loop box (see fig.\ref{fig11a} D) ) can be evaluated by directly integrating the Feynman Parameters via the HyperInt Maple library \cite{PanzerHyperInt}. The result is given by
\begin{eqnarray}
a^{(3)}(x)&=&\frac{\pi^2}{2}H_{-2,-2}-\frac{\pi^2(-15+2\pi^2)}{15}H_{-2}-\pi^2H_{-1}
+2\pi^2H_{-3,-1}-\frac{2\pi^2}{3}H_{-3,0}-\frac{2\pi^2}{3}H_{0,0}+\nonumber\\
&+&\frac{6-\pi^2}{3}H_{-2,0,0}-2H_{-1,0,0}+4H_{-3,-1,0,0}+H_{-2,-2,0,0}+
\frac{\pi^2-12\zeta_3}{3}H_{0}+4\zeta_3H_{-3}-\nonumber\\
&-&2\zeta_3H_{-2,0}+\frac{-504\pi^4+31\pi^6+3780\zeta_3(-1+2\zeta_3)}{1890},\nonumber\\
b^{(3)}(x)&=&-4\pi^2H_{-2,-1}+\frac{4\pi^2}{3}H_{-2,0}-\pi^2H_{-1,-2}+3\pi^2H_{-1,-1}+
\frac{2\pi^2}{3}H_{-1,0,0}-8H_{-2,-1,0,0}-\nonumber\\
&-&2H_{-1,-2,0,0}+6H_{-1,-1,0,0}-(\pi^2-4\zeta_3)H_{-1,0}-8\zeta_3H_{-2}
+\frac{4\pi^2+90\zeta_3}{15}H_{-1}.
\end{eqnarray}
The result for the four-loop box can also be found in \cite{PanzerHyperInt}.

In the main text, the expansion in powers of $y$ around the $z=-1+y$ point was investigated extensively. 
Here we will present a collection of various expansions of the $u$-channel box functions ($x=u/s$) in powers of $\delta=t/s$, which can be obtained from the
answers listed above by the HPL Mathematica library \cite{HPLtools}. All results for real parts presented below were also derived
from the closed expression for the four-point amplitude (\ref{AJseriesRep}) by the method described in section \ref{s4}. Since $z=1+2u/s=-1+2t/s$, delta parameter here
and $y$ are related as $y=-2\delta$. So at the one-loop level:
\begin{eqnarray}
\mbox{Re}\mathcal{B}^{(1)}&=&0-\frac{1}{2}\delta-\frac{1}{2}\delta^2-\frac{11}{24}\delta^3+O(\delta^4),\nonumber\\
\mbox{Im}\mathcal{B}^{(1)}&=&\pi+\frac{\pi}{2}\delta+\frac{\pi}{3}\delta^2
+\frac{\pi}{4}\delta^3+O(\delta^4).
\end{eqnarray}
At the two-loop level:
\begin{eqnarray}
\mbox{Re}\mathcal{B}^{(2)}&=&\left(\frac{-\pi^2}{3}+\frac{11\pi^4}{180}\right)+
\left(-\frac{1}{2}-\frac{\pi^2}{2}+\frac{11}{180}\pi^4\right)\delta
+O(\delta^2),\nonumber\\
\mbox{Im}\mathcal{B}^{(2)}&=&\left(2\pi-2\pi\zeta_3\right)+
\left(\frac{5\pi}{2}-2\pi\zeta_3\right)\delta+O(\delta^2).
\end{eqnarray}
And finally at  the three-loop level
\begin{eqnarray}
\mbox{Re}\mathcal{B}^{(3)}&=&-\frac{\pi^2}{3}+\frac{11\pi^4}{60}-6\zeta_3-\pi^2\zeta_3
-\frac{15}{2}\zeta_5+O(\delta),\nonumber\\
\mbox{Im}\mathcal{B}^{(3)}&=&2\pi+\frac{\pi^2}{3}+\frac{7\pi^5}{360}+6\pi\zeta_3+O(\delta).
\end{eqnarray}

In the main text, the Regge limit of large $z$ or equivalently large $x$ was also investigated. Using the results above, the following expansion of the box $\mathcal{B}^{(l)}(x)$ functions can be written ($\log(z)\sim\log(x)\equiv L$):
\begin{eqnarray}
x~\mathcal{B}^{(1)}&=&\underbrace{\frac{1}{2}L^2}_{LLA}+....
\end{eqnarray}
\begin{eqnarray}
x~\mathcal{B}^{(2)}&=&\underbrace{\frac{1}{12}L^4}_{LLA}+\underbrace{\frac{1}{3}L^3}_{NLA}
+\underbrace{\frac{\pi^2}{3}L^2}_{NNLA}
+\underbrace{\left(\frac{2\pi^2}{3}-2\zeta_3\right)L}_{NNNLA}+...
\end{eqnarray}
\begin{eqnarray}
x~\mathcal{B}^{(3)}&=&\underbrace{\frac{1}{144}L^6}_{LLA}+
\underbrace{\left(\frac{1}{12}+\frac{\pi^2}{3}\right)L^5}_{NLA}+
\underbrace{\left(\frac{1}{3}-\frac{\pi^2}{16}\right)L^4}_{NNLA}+
\underbrace{\left(\frac{1}{3}+
\frac{\pi^2}{2}-\frac{\zeta_3}{3}\right)L^3}_{NNNLA}
+\nonumber\\
&+&\left(\frac{7\pi^2}{6}+\frac{71\pi^4}{720}-2\zeta_3\right)L^2+
\left(\frac{2\pi^2}{3}+\frac{71\pi^4}{180}-2\zeta_3-\pi^2\zeta_3\right)L+...
\end{eqnarray}

\section{Integral representation of $\Omega_{\nu,J}(z)$ and Lerch Zeta function}\label{a2}

To obtain a representation of $\Omega_{\nu,J}(z)$, one should use the integral representation of Legendre polynomials of the following form:
\begin{equation}\label{LegendrePint}
P_k(z) = \frac{1}{2 \pi i} \int\limits_{[1,z]} \frac{(t^2-1)^k}{2^k(t-z)^{k+1}}dt,
\end{equation}
where the contour $[1,z]$ encircles the points $z$ and $1$ and does not contain or cross a half-interval ($-\infty,-1$]. 

Using this representation, one can consider $\Omega_{\nu,J}(z)$ written as
\begin{equation}
\Omega_{\nu,J}(z)=\frac{2^J}{\pi^2}\sinh^2(\pi\nu+i\pi J/2)\sum_{k=0}^J\frac{P_k(z)P_{J-k}(z)}{(J/2-k)^2+\nu^2},
\end{equation}
and  rewrite the sum over $k$ in the expression above in the following way:
\begin{equation}
\begin{gathered}
\sum_{k=0}^{J}\frac{P_k(z)P_{J-k}(z)}{(J/2-k)^2+\nu^2}=\frac{1}{(2\pi i)^2} \int\limits_{[1,z]} \int\limits_{[1,z]} dt_1 dt_2 \frac{(t_2^2-1)^J}{2^J(t_1-z)(t_2-z)^{J+1}} \times \\
    \times \sum_{k=0}^{J}\frac{[(t_1^2-1)(t_2-z)]^k}{((J/2-k)^2+\nu^2)[(t_2^2-1)(t_1-z)]^k}.
\end{gathered}
\end{equation}
Introducing the notation
\begin{equation}
 \mathcal{Z}\equiv\frac{(t_1^2-1)(t_2-z)}{(t_2^2-1)(t_1-z)},
\end{equation}    
we can evaluate the sum $\Sigma$ in terms of the so-called \emph{Lerch transcendent zeta functions} $\Phi(z,a,\alpha)$ \cite{LerchBook} (or Hurwitz-Lerch $\Phi$):
\begin{equation}
\begin{gathered}
    \Sigma\left(\mathcal{Z},\nu,J\right)\equiv\sum_{k=0}^{J}\frac{\mathcal{Z}^k}{(J/2-k)^2+\nu^2}=-\frac{i}{2 \nu} \left( \Phi\left(\mathcal{Z} ,1,-\frac{J}{2} - i \nu\right)-\Phi\left(\mathcal{Z} ,1,-\frac{J}{2} + i \nu\right)+\right. \\
    \left.+\mathcal{Z}^{1+J}\left(\Phi\left(\mathcal{Z} ,1,1+\frac{J}{2} + i \nu\right) - \Phi\left(\mathcal{Z} ,1,1+\frac{J}{2} - i \nu\right)\right) \right)
\label{sum to Lerch}
\end{gathered}
\end{equation}
This in turn allows one to obtain the representation (\ref{OmegaLerch}) for $\Omega_{\nu,J}(z)$ from the main text.

\par
Let us briefly discuss the definition and main properties of the Lerch transcendent zeta functions. The Lerch transcendent, can also be defined as the following series:
\begin{equation}
    \Phi(z,s,a)=\sum\limits_{n=0}^{\infty} \frac{z^n}{(n+a)^s},
\end{equation}
where: $a\neq0,-1,-2,\ldots$, $|z|<1$ or $|z|=1$ and $\mbox{Re}\:s >1$. 
Other values of $z$ can be also considered by means of analytical continuation. This can be done via contour integral of the following form:
\begin{equation}\label{LerchIntRepApp}
    \Phi(z,s,a)=\frac{1}{\Gamma(s)}\int^{+\infty}_{0}dt\frac{t^{s-1}\exp(-at)}{1-z\exp(-t)},
\end{equation}
where $\mbox{Re}\: s>0$, $\mbox{Re}\: a>0$ and $z\in\mathbb{C}/[1,+\infty)$. 

In some special cases the Lerch transcendent can be reduced to other special functions. Namely, \\
a) the polylogarithm is a special case of the Lerch transcendent Zeta given by 
\begin{equation}
    z \: \Phi(z,s,1) = \text{Li}_{s}(z),
\end{equation}
b) the Riemann zeta function is given by
\begin{equation}
    \Phi(1,s,1) = \zeta(z).
\end{equation}
c) the PolyGamma function is given by
\begin{equation}
    \Phi(1,n+1,z) = \frac{(-1)^{s+1}}{\Gamma(n+1)}\Psi^{(n)}(z).
\end{equation}

Using the integral representation (\ref{LerchIntRepApp}), one can prove \cite{LerchInv} the following relation between the Lerch Zeta function and that of the inverse argument:
\begin{equation}\label{LerchInvers}
    \Phi\left(z, 1, a\right)= z^{-1}\Phi\left(z^{-1}, 1,1-a\right)+\pi z^{-a}\left(\cot{\pi a}- i \: \text{sgn}(\varphi)\right),
\end{equation}
where $\varphi=\arg(-\ln{z})$.

There exist several asymptotic formulas for the Lerch Zeta function. One of these relations is given by:
\begin{equation}\label{LerchSeriesFull}
    \Phi(z, s, a) = \frac{1}{1-z}\frac{1}{a^s} +\sum_{n=1}^{N-1}\frac{(-1)^n\text{Li}_{-n}(z)}{n!}\frac{(s)_n}{a^{n+s}}+ O(a^{-N+s}),
\end{equation}
where $\mbox{Arg} \: a <\pi$, $s \in \mathbb{C}$, $z \in \mathcal{C}_a$, $\mathcal{C}_a=\mathbb{C}/[1,+\infty)$ if $\mbox{Re}\: a>0$ or $\mathcal{C}_a=|z|<1$ if $\mbox{Re}\: a<0$ and $(s)_n$ is the Pochhammer symbol.
In our case, we use a particular version of this relation:
\begin{equation}\label{LerchSeries1}
    \Phi(z, 1, a) = \frac{1}{1-z}\frac{1}{a} + O(a^{-2}).
\end{equation}

\section{Strong coupling supplementary}\label{a3}

Here we will give the details of the derivation of (\ref{AmplStrongCouplingFinal}) and (\ref{AmplStrongCouplingFinalz1}) from the main text. Our starting point will
be expression (\ref{AmplitudeStrongJint}) for the amplitude
\begin{equation}\label{AmplStr1}
    A_4^{D=4,u}=  8\pi \sum_{J \geq 0}^{\infty} \sum_{\nu=\nu_2(J), \nu_4(J)} \frac{2(J/g+1)(4\nu^2+(J/g+1)^2)\nu}{(4 \nu^2+(J/g+1)^2+1)}\frac{\Omega_{\nu,J/g}(z)}{2^{J/g}}.
\end{equation}
As was explained in the main text, the $\nu$ integral can be evaluated by residues. 
If we enclose the integration contour over a lower half of the complex plane, only the $\nu_2$ and $\nu_4$ roots will contribute. Using the representation for $\Omega_{\nu,J/g}(z)$ in the from (\ref{OmegaLerch}), we can try to take advantage of the presence of the large parameter $1/g$, expand the integrand in powers of $g$ and evaluate all the remaining integrals. 

Let us start with the $\nu=\nu_4(J)$ contribution to the amplitude and consider the expansion of $\Omega_{\nu,J/g}(z)$ first.
When we substitute
$$
\nu=\nu_4(J) = \frac{-i}{2g} \sqrt{2 g^2 + 2 g J + J^2 + 2 \sqrt{4 + g^4 + 2 g^3 J + g^2 J^2}},
$$
in (\ref{AmplStr1}) the real parts of the third Lerch Zeta function arguments in $\Omega_{\nu,J/g}(z)$ are Re$(J/2g+i \nu) > 0$, and Re$(J/2g-i \nu) < 0$.
If we want to use asymptotic expression (\ref{LerchSeries1}) for $\Omega_{\nu,J/g}(z)$, we should transform the first and fourth Lerch Zeta functions according to (\ref{LerchInvers}). After such transformation all four Lerch Zeta functions contributing to $\Omega_{\nu,J/g}(z)$ will have their third argument real parts $>0$ and we can use asymptotic formula (\ref{LerchSeries1}).

After the use of (\ref{LerchInvers}) and (\ref{LerchSeries1}), the $\Sigma(\mathcal{Z},\nu,J)$ part of $\Omega_{\nu,J/g}(z)$ takes the following form:
\begin{equation}
\begin{gathered}
    \Sigma(\mathcal{Z},\nu,J)\Big{|}_{\nu=\nu_4} = -\frac{i}{2 \nu} \Big{(} \frac{\mathcal{Z}^{1+J}-1}{1 - \mathcal{Z}}\left[(1+J/2+i\nu)^{-1}+(-J/2+i\nu)^{-1}\right] + \\
    + \pi \mathcal{Z}^{J/2+i\nu} \left[\cot{\pi(-J/2+i\nu)}-\cot{\pi(J/2+i\nu)} \right] \Big{)} \Big{|}_{\nu=\nu_4(J)},
    \label{Sum Hurwitz final form}
\end{gathered}
\end{equation}
which can be further expanded in powers of $g$:
\begin{equation}
\begin{gathered}
     \Sigma(\mathcal{Z},\nu,J)\Big{|}_{\nu=\nu_4}= \frac{g}{\sqrt{4+J^2}} \left( \frac{\mathcal{Z}^{1+J/g}-1}{1 - \mathcal{Z}} g \: \sqrt{4+J^2} + \right.
     \\
     + \pi \mathcal{Z}^{(J+\sqrt{4+J^2})/2g} \left\{\cot{\left[\pi\left(\frac{-J+\sqrt{4+J^2}}{2g}+\frac{J}{2\sqrt{4+J^2}}\right)\right]} \right.-
     \\
     \left.\left. -\cot{\left[\pi\left(\frac{J+\sqrt{4+J^2}}{2g}+\frac{J}{2\sqrt{4+J^2}}\right)\right]} \right\} \right)+\ldots . 
\end{gathered}
\end{equation}
Here we have neglected all terms suppressed by powers of $g$.
It is convenient to separate $\Omega_{\nu,J}$ into two pars $\Omega=\Omega^{(a)}+\Omega^{(b)}$. 
Then taking into account the asymptotic expression for $\Sigma(\mathcal{Z},\nu,J)$, we can obtain:
\begin{equation}
\begin{gathered}
    \Omega^{(a)}_{\nu_4,J/g}(z) = \frac{g^2}{\pi^2}\sin^2\left(\pi\left[\frac{-J+\sqrt{4+J^2}}{2g}+\frac{J}{2\sqrt{4+J^2}}\right]\right)  \times\\
    \times \frac{1}{(2\pi i)^2} \int\limits_{[1,z]} \int\limits_{[1,z]} dt_1 dt_2 \frac{(t_2^2-1)^{J/g} (\mathcal{Z}^{1+J/g}-1)}{(t_1-z)(t_2-z)^{J/g+1}(\mathcal{Z}-1)}
    \label{OmegaNU4 1-st part}
\end{gathered}
\end{equation}
and
\begin{equation}
\begin{gathered}
    \Omega^{(b)}_{\nu_4,J/g}(z)=\frac{-g}{\pi \sqrt{4+J^2}}\sin\left(\pi\left[\frac{-J+\sqrt{4+J^2}}{2g}+\frac{J}{2\sqrt{4+J^2}}\right]\right)  \times \\
    \times \left\{\frac{\sin{\frac{\pi J}{g}}}{\sin{\left[\pi\left(\frac{J+\sqrt{4+J^2}}{2g}+\frac{J}{2\sqrt{4+J^2}}\right)\right]}} 
    \right\} 
    \cdot \frac{1}{(2\pi i)^2} \int\limits_{[1,z]} \int\limits_{[1,z]} dt_1 dt_2 \frac{(t_2^2-1)^{J/g}   \mathcal{Z}^{\frac{J+\sqrt{4+J^2}}{2g}}}{(t_1-z)(t_2-z)^{J/g+1}}
    \label{OmegaNU4 2-nd part}
\end{gathered}
\end{equation}
Let us additionally simplify $\Omega^{(a)}_{\nu_4,J/g}(z)$.
One can note that the double integral of (\ref{OmegaNU4 1-st part}) is given by the
Chebyshev polynomial of the second kind:
\begin{equation}
\begin{gathered}
   2^{J/g} U_{J/g}(z) = \frac{1}{(2\pi i)^2} \int\limits_{[1,z]} \int\limits_{[1,z]} dt_1 dt_2 \frac{(t_2^2-1)^{J/g} (\mathcal{Z}^{1+J/g}-1)}{(t_1-z)(t_2-z)^{J/g+1}(\mathcal{Z}-1)}.
    \label{Chebyshev}
\end{gathered}
\end{equation}
One can see that this is indeed the case by rewriting $U_{J}(z)$ as the sum of products of the
Legendre polynomials
$$
U_{J}(z)=\sum_{k=0}^J P_{k}(z)P_{J-k}(z),
$$
and using the integral representation (\ref{LegendrePint}) for each of $P_{J}(z)$.
This allows us to write
\begin{equation}
    \Omega^{(a)}_{\nu_2,J/g}(z) = 2^{J/g}\frac{g^2}{\pi^2}\sin^2\left(\pi\left[\frac{J+\sqrt{4+J^2}}{2g}+\frac{J}{2\sqrt{4+J^2}}\right]\right) \: U_{J/g}(z) 
    \label{OmegaNU2 1-st part, new}
\end{equation}
The rest of the amplitude evaluated in $\nu = \nu_4$ in the strong coupling limit takes the form:
\begin{equation}
\begin{gathered}
   \left. \frac{2^{1-J/g}(J/g+1)(4\nu^2+(J/g+1)^2)\nu}{(4 \nu^2+(J/g+1)^2+1)} \right|_{\nu=\nu_4(J)}=-\frac{i 2^{-J/g} J \sqrt{4 + J^2}}{g^2}+\ldots ,
\end{gathered}
\end{equation}
where we once again neglected all positive powers of $g$.
The latter, with the combination with the results for $\Omega^{(a)}_{\nu_4,J/g}(z)$ and $\Omega^{(b)}_{\nu_4,J/g}(z)$, allows us to write the contribution to the amplitude from the $\nu = \nu_4$ pole as:
\begin{equation}
\begin{gathered}
    A^{D=4}_4 |_{\nu=\nu_4} = -\frac{4}{\pi}  \int\limits_{C}\frac{d J}{\sin{(\pi J/g)}} J \sqrt{4 + J^2} \: U_{J/g}(z) \times \\
    \times \sin^2\left(\pi\left[\frac{-J+\sqrt{4+J^2}}{2g}+\frac{J}{2\sqrt{4+J^2}}\right]\right)+\ldots 
\end{gathered}
\end{equation}
where we have neglected all terms containing positive powers of $g$ and/or contributions containing the $g\times\sin^1 (\ldots)$ factor. Note that the latter contributions are generated only from the $\Omega^{(b)}_{\nu_4,J/g}(z)$ term.
\\

Let us now consider the contribution from the $\nu=\nu_2(J)$ pole to the amplitude.
\begin{equation}
    \nu_2(J) = -\frac{1}{2g}\sqrt{-2 g^2 - 2 g j - j^2 + 2 \sqrt{4 + g^4 + 2 g^3 j + g^2 j^2}}
\end{equation}
As in the previous case, let us consider the $\Omega_{\nu,J/g}(z)$ function first.
In this case Re$(J/2g+ \\ + i \nu_2(J)) > 0$ and Re$(J/2g-i \nu_2(J)) > 0$, so following similar steps as 
before and transforming the first and second Lerch functions in $\Sigma(\mathcal{Z},\nu,J)$, we obtain:
\begin{equation}
\begin{gathered}
     \Sigma(\mathcal{Z},\nu,J)\Big{|}_{\nu=\nu_2} = -\frac{i}{2 \nu} \Big{(} \frac{\mathcal{Z}^{1+J}-1}{1 - \mathcal{Z}}\left[w^{-1}-\bar{w}^{-1}\right] + \\
     + \pi \left[ \mathcal{Z}^{\bar{w}-1} \cot{\pi(\bar{w}-1)}-\mathcal{Z}^{w-1}\cot{\pi(w-1)} \right] +i \pi \: \text{sng}(\varphi)[\mathcal{Z}^{w-1}-\mathcal{Z}^{\bar{w}-1}] \Big{)} \Big{|}_{\nu=\nu_2(J)},
     \label{sum to Hurwitz nu2 asimpt.}
\end{gathered}
\end{equation}
where we have introduced the notation $w \equiv 1+j/2+i\nu$, $\bar{w} \equiv 1+j/2-i\nu$. 
This expression can be further expanded in powers of $g$, which gives us:
\begin{equation}
\begin{gathered}
\Sigma(\mathcal{Z},\nu,J)\Big{|}_{\nu=\nu_2}= \frac{i g}{\sqrt{4-J^2}} \left( \frac{\mathcal{Z}^{1+J/g}-1}{1 - \mathcal{Z}} ig \sqrt{4-J^2}+ \right.
\\
+ \pi \left\{ \mathcal{Z}^{(J+i\sqrt{4-J^2})/2g} \cot{\left[\pi\left(\frac{J+i\sqrt{4-J^2}}{2g}-\frac{iJ}{2\sqrt{4-J^2}}\right)\right]}- \right.\\
\left.-\mathcal{Z}^{(J-i\sqrt{4-J^2})/2g} \cot{\left[\pi\left(\frac{J-i\sqrt{4-J^2}}{2g}+\frac{iJ}{2\sqrt{4-J^2}}\right)\right]} \right\} +\\
+i \pi \: \text{sng}(\varphi)[\mathcal{Z}^{(J-i\sqrt{4-J^2})/2g}-\mathcal{Z}^{(J+i\sqrt{4-J^2})/2g}] \Big{)}+\ldots. 
\end{gathered}
\end{equation}
As before, let us separate $\Omega_{\nu,J/g}(z)$ into two parts $\Omega=\Omega^{(a)}+\Omega^{(b)}$, with:
\begin{equation}
\begin{gathered}
    \Omega^{(a)}_{\nu_2,J/g}(z) = -\frac{g^2}{\pi^2}\sin^2\left(\pi\left[\frac{J+i \sqrt{4-J^2}}{2g}-\frac{iJ}{2\sqrt{4-J^2}}\right]\right)  \times\\
    \times \frac{1}{(2\pi i)^2} \int\limits_{[1,z]} \int\limits_{[1,z]} dt_1 dt_2 \frac{(t_2^2-1)^{J/g} (\mathcal{Z}^{1+J/g}-1)}{(t_1-z)(t_2-z)^{J/g+1}(\mathcal{Z}-1)}
    \label{OmegaNU2 1-st part}
\end{gathered}
\end{equation}
and
\begin{equation*}
\begin{gathered}
    \Omega^{(b)}_{\nu_2,J/g}(z)=\frac{-i g}{\pi \sqrt{4-J^2}}\sin^2\left(\pi\left[\frac{J+i \sqrt{4-J^2}}{2g}-\frac{iJ}{2\sqrt{4-J^2}}\right]\right)  \times \\
    \times \left\{\cot{\left[\pi\left(\frac{J+i\sqrt{4-J^2}}{2g}-\frac{iJ}{2\sqrt{4-J^2}}\right)\right]}
    \cdot \frac{1}{(2\pi i)^2} \int\limits_{[1,z]} \int\limits_{[1,z]} dt_1 dt_2 \frac{(t_2^2-1)^{J/g}   \mathcal{Z}^{\frac{J+i\sqrt{4-J^2}}{2g}}}{(t_1-z)(t_2-z)^{J/g+1}}-\right.
    \\
    \left.-\cot{\left[\pi\left(\frac{J-i\sqrt{4-J^2}}{2g}+\frac{iJ}{2\sqrt{4-J^2}}\right)\right]} \cdot \frac{1}{(2\pi i)^2} \int\limits_{[1,z]} \int\limits_{[1,z]} dt_1 dt_2 \frac{(t_2^2-1)^{J/g}   \mathcal{Z}^{\frac{J-i\sqrt{4-J^2}}{2g}}}{(t_1-z)(t_2-z)^{J/g+1}} + \right.
    \\
    \left. +i \: \text{sng}(\varphi) \cdot \frac{1}{(2\pi i)^2} \int\limits_{[1,z]} \int\limits_{[1,z]} dt_1 dt_2 \frac{(t_2^2-1)^{J/g}   [\mathcal{Z}^{(J-i\sqrt{4-J^2})/2g}-\mathcal{Z}^{(J+i\sqrt{4-J^2})/2g}]}{(t_1-z)(t_2-z)^{J/g+1}} \right\}.
\end{gathered}
\end{equation*}
Using identical manipulations as in the previous case, $\Omega^{(a)}$ can be simplified into:
\begin{equation}
\begin{gathered}
     \Omega^{(a)}_{\nu_2,J/g}(z) = g^2 \frac{2^{J/g}}{\pi^2}\sin^2\left(\pi\left[\frac{J+i \sqrt{4-J^2}}{2g}-\frac{iJ}{2\sqrt{4-J^2}}\right]\right)   \times \\
    \times U_{J/g}(z)
    \label{OmegaNU2 1-st final}
\end{gathered}
\end{equation}
All other than $\Omega_{\nu,J/g}(z)$ terms in the amplitude can be expanded at $\nu = \nu_2$
in powers of g as:
\begin{equation}
\begin{gathered}
   \left. \frac{2^{1-J/g}(J/g+1)(4\nu^2+(J/g+1)^2)\nu}{(4 \nu^2+(J/g+1)^2+1)} \right|_{\nu=\nu_2(J)}=-\frac{ 2^{-J/g} J \sqrt{4 - J^2}}{g^2}+\ldots.
\end{gathered}
\end{equation}
Combining this result with the expressions for $\Omega_{\nu,J/g}^{(a)}(z)$ and $\Omega_{\nu,J/g}^{(b)}(z)$,
we can finally  obtain:
\begin{equation}
\begin{gathered}
    A^{D=4,u}_4 |_{\nu=\nu_2} = \frac{-8 \pi}{\pi^2} \sum\limits_{J=0}^{\infty} J \sqrt{4 - J^2} \: U_{J/g}(z) \times \\
    \times \sin^2\left(\pi\left[\frac{J+i\sqrt{4-J^2}}{2g}-\frac{i J}{2\sqrt{4-J^2}}\right]\right)+\ldots.
\end{gathered}
\end{equation}
Transforming the overall $\sin$ factor in this expression according to:
\begin{equation}
    \sin^2\left(\pi\left[\frac{J+i\sqrt{4-J^2}}{2g}-\frac{i J}{2\sqrt{4-J^2}}\right]\right) \approx -\frac{1}{4} \exp{\left(\frac{-i \pi J}{g} + \frac{\pi \sqrt{4-J^2}}{g}\right)}
    \label{approx}
\end{equation}
and rewriting $U_{J/g}(z)$ as:
\begin{equation}
    U_{J/g}(z) \approx \frac{\exp{\left((1+\frac{J}{g})\mathbf{L}\right)}}{2\sqrt{z^2-1}},
\end{equation}
which is valid for $z>1$ and where $\mathbf{L} \equiv \log{(z+\sqrt{z^2-1})}$, we finally arrive at\\
\begin{equation}\label{Aleading}
\begin{gathered}
    A^{D=4,u}_4 |_{\nu=\nu_2} = \frac{1}{2\pi i (\sqrt{z^2-1})}  \int\limits_{C}\frac{d (J/g)}{\sin{(\pi J/g)}} J \sqrt{4 - J^2} \exp{\left(\frac{\pi \sqrt{4-J^2}}{g}+(1+\frac{J}{g})\mathbf{L}\right)}+\ldots.
\end{gathered}    
\end{equation}
where, as in the previous case, we have neglected all terms containing positive powers of the $g$ and/or
$g\times \sin^1$ factors. 

Comparing the $A^{D=4,u}_4 |_{\nu=\nu_4}$ and $A^{D=4,u}_4 |_{\nu=\nu_2}$ contributions to the amplitude, we conclude that the leading (exponentially enhanced $\sim \exp(1/g)$) strong coupling contribution will be given by the $A^{D=4,u}_4 |_{\nu=\nu_2}$ part listed in (\ref{Aleading}) only. All other terms will generate subleading contributions. In (\ref{Aleading}) the $J$ integration can be evaluated by the steepest descent method. 

Let us note that ($\ref{approx}$) is valid for
\begin{equation}
    \Big{|}\frac{\sqrt{4-J_0^2}}{g} \Big{|} \gg \Big{|}\frac{-J_0}{\sqrt{4-J_0^2}}\Big{|},
\end{equation}
which can be violated in the vicinity of the steepest descend point $J_0 \sim 2$ which is equivalent to a sufficiently large $z$. One can show that the condition on $z$ takes the form:
\begin{equation}
    L \ll \frac{\sqrt{2}\pi}{\sqrt{g}} \quad \Rightarrow \quad z \ll \text{e}^{\sqrt{2}\pi/\sqrt{g}}.
\end{equation}

The point $z=1$ should be considered separately. Substituting $z \rightarrow 1$ in the integrand of (\ref{OmegaNU2 1-st part}), we can obtain:
\begin{equation}
    \frac{1}{(2\pi i)^2} \int\limits_{[1,z]} \int\limits_{[1,z]} dt_1 dt_2 \frac{(t_2^2-1)^{J/g} (\mathcal{Z}^{1+J/g}-1)}{(t_1-z)(t_2-z)^{J/g+1}(\mathcal{Z}-1)} \Big{|}_{z \rightarrow 1}= 2^{J/g}(1+J/g).
\end{equation}
Then $\Omega^{(a)}_{\nu_2,J/g}(z=1)$  takes the form:
\begin{equation}
     \Omega^{(1)}_{\nu_2,J/g}(z=1) = g^2 \frac{2^{J/g}}{\pi^2}\sin^2\left(\pi\left[\frac{J+i \sqrt{4-J^2}}{2g}-\frac{iJ}{2\sqrt{4-J^2}}\right]\right)\left(1+\frac{J}{g}\right).
    \label{OmegaNU2 1-st final, z=1}
\end{equation}
All other terms except $\Omega^{(a)}_{\nu_2,J/g}$ remain the same which gives us the following approximate expression for the amplitude
\begin{equation}
\begin{gathered}
     A^{D=4,u}_4|_{\nu=\nu_2,z=1} = \frac{1}{\pi i}  \int\limits_{C}\frac{d (J/g)}{\sin{(\pi J/g)}} J \sqrt{4-J^2} (1+J/g) \exp{\left(\frac{\pi \sqrt{4-J^2}}{g}\right)}.
\end{gathered}    
\end{equation}
Evaluating the $J$ integral by the steepest descend method  around $J_0 = 0$, we obtain:
\begin{equation}
    A^{D=4,u}_4 |_{\nu=\nu_2,z=1} = \sqrt{g} \frac{ 8\pi}{2\pi^3 i}  \exp{\left(2 \pi/g\right)}.
\end{equation}

\section{$\chi$-sector and corresponding fixed point}\label{a4}
Here we will consider double-trace four-point colour ordered amplitude for the $\chi$-sector. We will examine the UV finiteness condition for this amplitude and will find a fixed point for the fishnet theory considered in the main text up to two loops in PT.

In \cite{FishnetGeneralD} it was pointed out that in $D$ dimension the double-trace part of Lagrangian
should be given by:
\begin{eqnarray}
\mathcal{L}_{c.t.}^{(6,1)}/(4\pi)^{D/2}&=&\alpha_1(g)
\mbox{tr}\left(\phi\phi\right)\mbox{tr}\left(\phi^*\phi^*\right)+
\alpha_1(g)
\mbox{tr}\left(\chi\chi\right)\mbox{tr}\left(\chi^*\chi^*\right)+\nonumber\\
&=&g\left(
\mbox{tr}\left(\phi\chi\right)\mbox{tr}\left(\phi^*\chi^*\right)
+\mbox{tr}\left(\phi^*\chi\right)\mbox{tr}\left(\phi\chi^*\right)
\right),
\end{eqnarray}
and that in the $\omega \neq D/4$ "inhomogeneous" case the $\alpha_1$ part should be absent because
$\alpha_1$ becomes dimensionful, which contradicts the conformal invariance of the theory. We see that indeed we do not need the $\mbox{tr}\left(\phi\phi\right)\mbox{tr}\left(\phi^*\phi^*\right)$ term due to the UV finiteness
of the $\phi$-sector at the level of four-point amplitudes, but the amplitudes in the $\chi$-sector will likely require
non-trivial double-trace interactions for the theory to be conformal.
The contribution to the $\chi$-amplitude in the first order of PT from the single-trace part of Lagrangian (\ref{D6LagrMain}) is given by a pair of bubble diagrams (see fig. \ref{fig4} where one has to inverse
the dot positions to obtain diagrams for the $\chi$-sector): 
\begin{equation}\label{chiSingleTraceAmpl}
    A_{s.t.}=g^4 (\upsilon(t)+\upsilon(u))+\ldots
\end{equation}
where
\begin{equation}
    \upsilon(s_{jk})= \int \frac{d^6 q}{(q+K)^2 q^2}= \frac{ \Gamma(2-\epsilon)^2 \Gamma(-1+\epsilon)}{\Gamma(4-2\epsilon)}(-s_{jk})^{1-\epsilon}.
\end{equation}
This contribution is UV divergent and at the one-loop level the singular part of $A_{s.t.}$ is given by
\begin{equation}
   \mbox{sing.}~A_{s.t.}=g^4\frac{t+u}{6 \epsilon}.
\end{equation}
We suggest a resolution to this problem by adding the double-trace $\chi$-interaction terms with the 
derivatives $\sim\partial^2~\mbox{tr}(\chi\chi)\mbox{tr}(\chi^*\chi^*)$.
This will allow one to obtain a dimensionless coupling constant before these terms. In general, there are three not independent, due to the full derivative constraint, ways to distribute a derivative between the $\chi$-fields. We choose them as:
\begin{eqnarray}
\alpha_1(g)\mbox{tr}\left(\chi\partial^2\chi\right)\mbox{tr}\left(\chi^*\chi^*\right)+
\alpha_2(g)\mbox{tr}\left(\chi\partial^{\mu}\chi\right)\mbox{tr}\left(\chi^*\partial_{\mu}\chi^*\right)
+c.c.
\end{eqnarray}

To find the fixed point solution $\alpha_i(g)$, one can consider the renormalization of two-point double-trace correlators similar to \cite{Grabner:2017pgm}. However, we would like to make a shortcut and will try to find the fixed point solution directly from the condition of UV-finiteness of the four-point $\chi$-sector amplitude with off-shell legs at least in lower orders of PT. Indeed the amplitude itself is given by the same diagrams as the correlation function but with pairs of external legs glued to the points. So we expect that the UV properties of both objects are closely related.
We also want to mention that the question of the existence of a common set of eigenfunctions between
this new modified $\mathcal{V}$ operator (i.e. double-trace $\chi$-operators with derivatives) and $\mathcal{H}_2$ should be separately investigated. We will however avoid this discussion in the current publication.

\begin{figure}[ht]
 \begin{center}
  \epsfxsize=8cm
 \epsffile{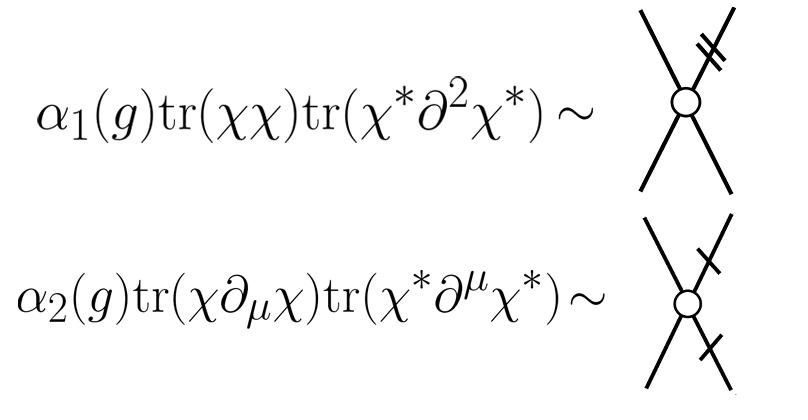}
 \end{center}
\caption{Vertex function of double-trace terms.} \label{appfigv1}
 \end{figure}

So to cancel the UV-pole in (\ref{chiSingleTraceAmpl}), one has to fine tune the constants $\alpha_{i}$ in the  double-trace part
of the Lagrangian:
\begin{equation}\label{weakexp}
    \alpha_i(g)=\alpha_{i,2} g^4+\alpha_{i,3} g^6+...+\alpha_{i,n}g^{2n}+...
\end{equation}
The coefficients $\alpha_{i,j}$  can then be found from the condition of cancellation of the UV poles in the amplitude with momenta $p_1^2=p^2_2=0$, $p_3^2\neq p_4^2\neq 0$.
The double-trace vertices are shown in fig. \ref{appfigv1}
where double notches on the line are the same squared momenta and pair-notches on two adjacent lines are the contracted different momenta. The one-loop $\chi$-field diagrams generated by these vertices have the following general structure:
\begin{equation}
  g^4\int d^6 q \frac{\text{N}_i(q,p_3,p_4)}{(q^2)^2((q+K)^2)^2},
\end{equation} 
where $\text{N}_i(q,p_3,p_4)$ stands for the numerator of each integral. 
It is convenient to separate all these diagrams into three subsets: generated by $(\partial^2)(\partial^2)$, generated by $(\partial\partial)(\partial\partial)$ and generated by $(\partial^2)(\partial\partial)$ vertices, respectively. 
\begin{figure}[ht]
 \begin{center}
  \epsfxsize=9cm
 \epsffile{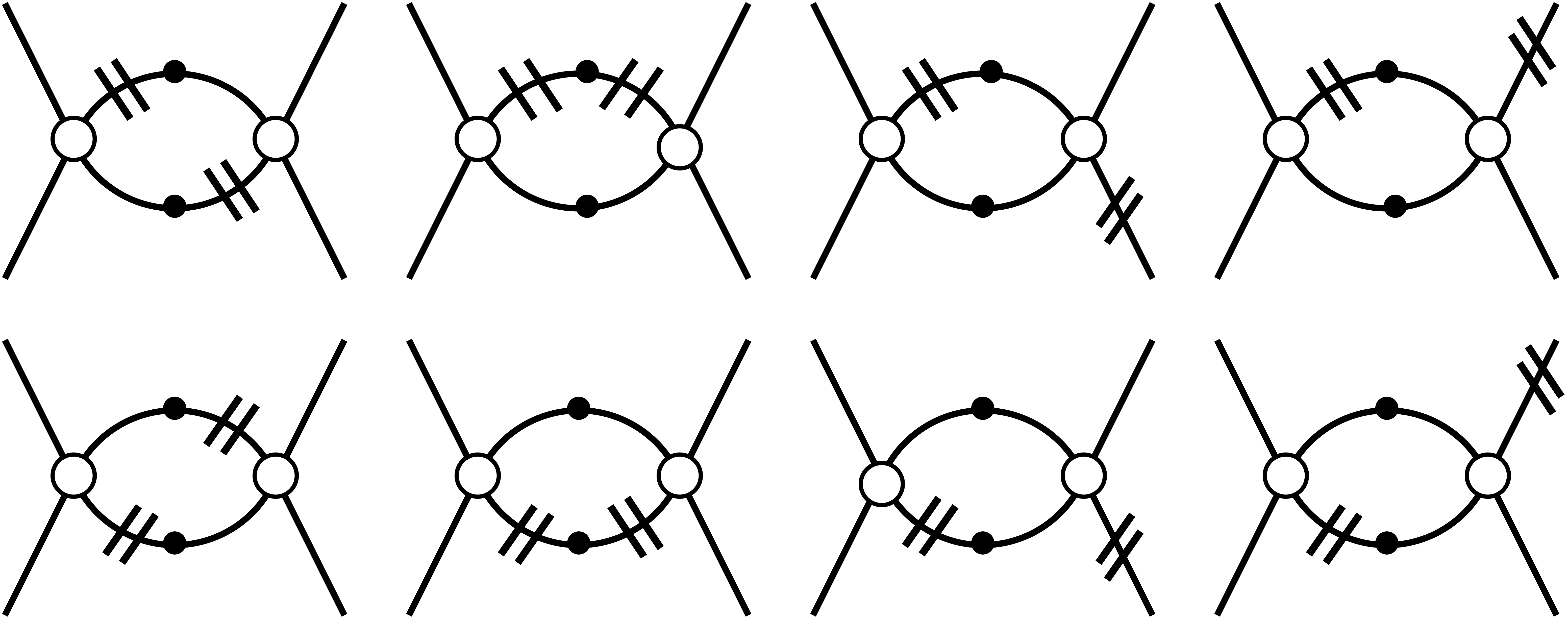}
 \end{center}
\caption{Set of 1-loop integrals from the double-trace Lagrangian contributing to the amplitude $A_{\partial^2 \partial^2}$.} \label{appfigd2}
 \end{figure}

In the case of the $(\partial^2)(\partial^2)$ contribution the corresponding Feynman diagrams are represented in fig.\ref{appfigd2} and the corresponding numerator is given by:
\begin{equation}
    \sum_i \text{N}_i=p_3^2 p_4^2+(p_3^2+p_4^2)(q^2 + Q^2)+ (q^2)^2+(Q^2)^2+q^2 Q^2,
\end{equation}
where $Q = (q+K)$.
The result is given by
\begin{equation}
\mbox{sing.}~A_{\partial^2 \partial^2}=\alpha_{1,2}^2 g^4 \frac{s-3 p_3^2-3 p_4^2}{3 \epsilon }.
\end{equation}

\begin{figure}[ht]
 \begin{center}
  \epsfxsize=9cm
 \epsffile{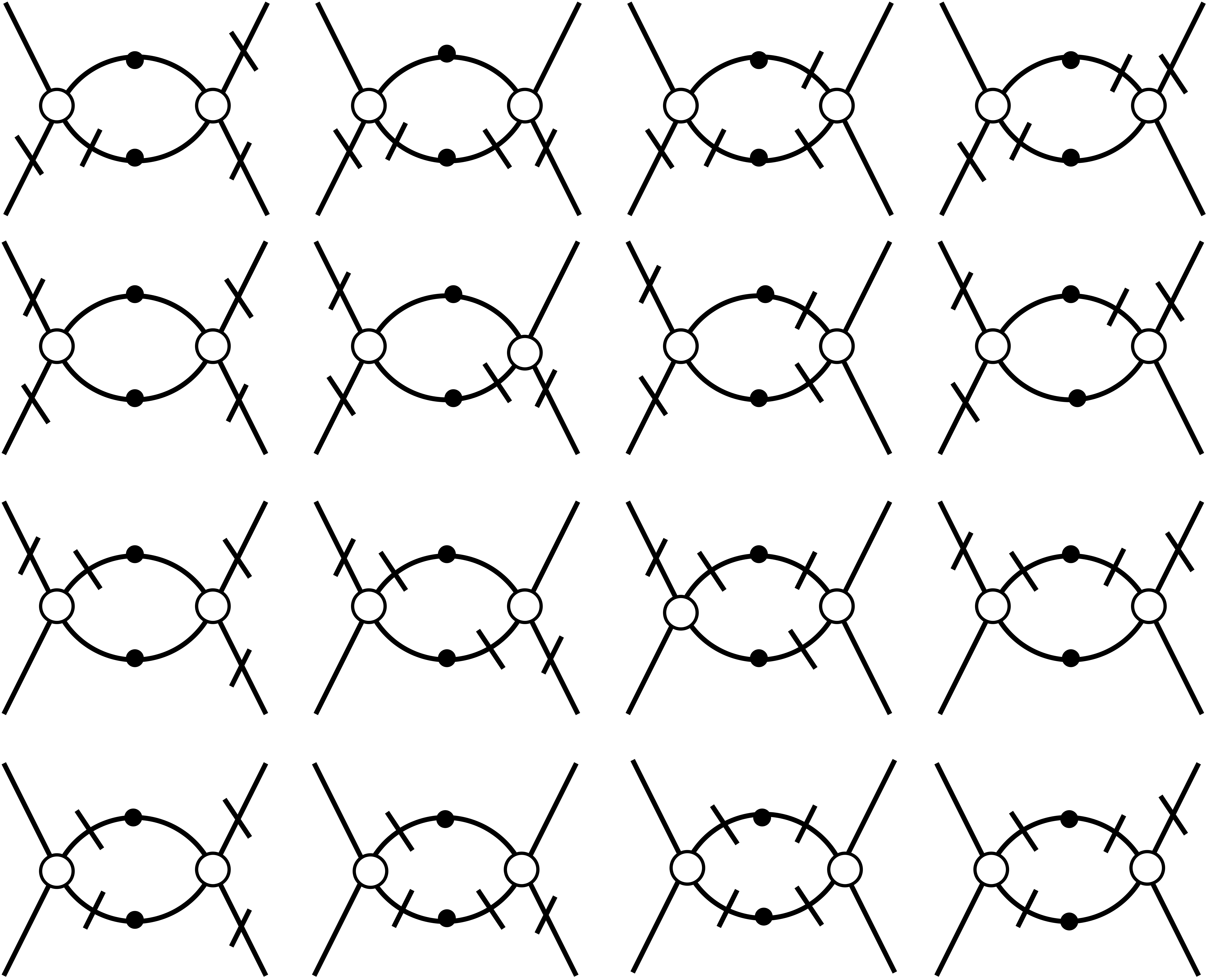}
 \end{center}
\caption{Set of 1-loop integrals from the double-trace Lagrangian contributing to the amplitude $A_{(\partial \partial) (\partial \partial)}$.} \label{appfigdddd}
 \end{figure}

In the case of the $(\partial\partial)(\partial\partial)$ diagrams the corresponding numerator is given by
\begin{multline}
  \sum_i N_i = (q p_3+q p_2+Q p_4+Qp_1+qQ) q Q + q p_3 q p_2+Q p_4 Q p_1+q p_3 Q p_1+\\+q p_2 Q p_4+p_1 p_2 p_3 p_4+qp_2 Qp_1+Q p_4 q p_3,
\end{multline}
and the diagrams themselves are represented in fig.\ref{appfigdddd}. 
This allows one to compute the singular part of the corresponding sum of diagrams:
\begin{equation}
  \mbox{sing.}~A_{(\partial \partial) (\partial \partial)}=\alpha_{2,2}^2 g^4 \frac{p_3^2+p_4^2+s}{4 \epsilon }
\end{equation}

\begin{figure}[ht]
 \begin{center}
  \epsfxsize=12cm
 \epsffile{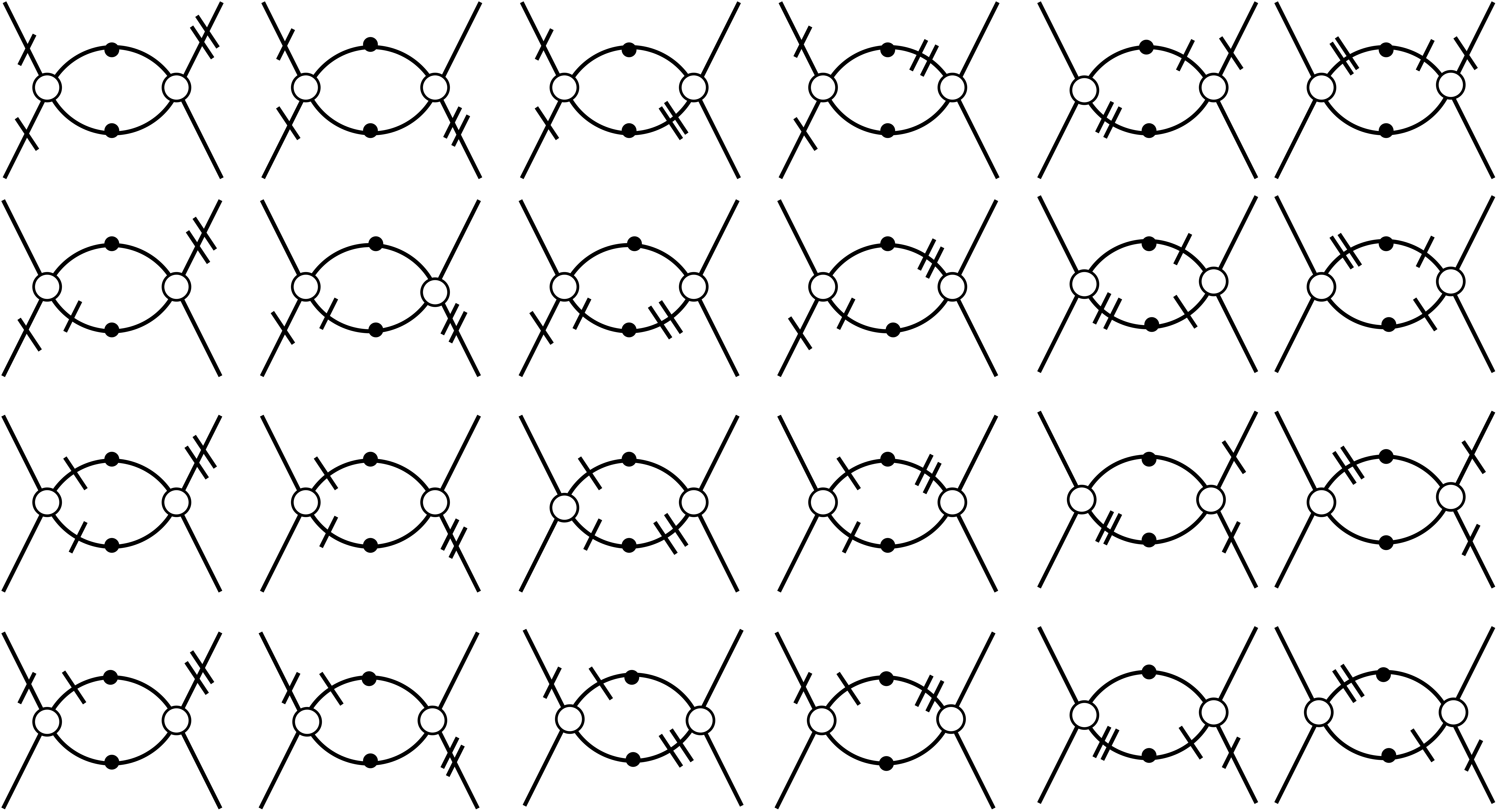}
 \end{center}
\caption{Set of 1-loop integrals from the double-trace Lagrangian contributing to the amplitude $A_{(\partial^2) (\partial \partial)}$. 
} \label{appfigddd2}
 \end{figure}

Finally, in the case of the $(\partial^2)(\partial\partial)$ diagrams the corresponding numerator is given by
\begin{multline}
    \sum_i N_i = (q p_2+q p_3+Q p_1+Q p_4+p_1 p_2+p_3 p_4)(Q^2+q^2)+\\+(p_3^2+p4^2) qQ+(q^2 + Q^2)qQ,
\end{multline}
and the diagrams themselves are represented in fig.\ref{appfigddd2}.
This allows one to compute the singular part of the corresponding sum of diagrams:
\begin{equation}
\mbox{sing.}~A_{\partial^2 (\partial\partial)} =\alpha_{1,2} \alpha_{2,2} g^4 \frac{5 s }{6 \epsilon }.
\end{equation}

Combining all contributions together, one can see that the singular part of the double-trace contribution to the amplitude is given by: 
\begin{multline}
\mbox{sing.}~A_{d.t.}=\mbox{sing.}\left(A_{\partial^2 (\partial\partial)}+ A_{(\partial \partial) (\partial \partial)}+A_{\partial^2 \partial^2}\right)=\frac{g^2}{12 \epsilon } \left(\left(4 \alpha _1^2+10 \alpha _2 \alpha _1+3 \alpha _2^2\right) s\right.  \\ \left. -3\left(4 \alpha _1^2-\alpha _2^2\right) \left(p_3^2+p_4^2\right) \right).
\end{multline}
Here we use the relation $s+t+u=p_3^2+p_4^2$ between the Mandelstam variables.
Taking into account the single-trace part of the amplitude, one can get:
\begin{equation}
\mbox{sing.}\left(A_{d.t.}+ A_{s.t.}\right)=\frac{g^4 }{6 \epsilon }(t+u)+\frac{g^2}{12 \epsilon } \left(\left(4 \alpha _1^2+10 \alpha _2 \alpha _1+3 \alpha _2^2\right) s -3\left(4 \alpha _1^2-\alpha _2^2\right) \left(p_3^2+p_4^2\right) \right).
\end{equation}
This allows one to obtain the fixed point condition for two parameters $\alpha_{1,2}$ and $\alpha_{2,2}$ at the one-loop order, requiring UV-cancellation: 
\begin{eqnarray}\label{zerocond}
 \left\{\begin{matrix}
3 \left(4 \alpha _1^2-\alpha _2^2\right)= 2\\ 
4 \alpha _1^2+10 \alpha _2 \alpha _1+3 \alpha _2^2= 2.
\end{matrix}\right.
\end{eqnarray}
Note that here there are two free parameters and two conditions that make the system uniquely solvable. This is where off shell momenta $p_3$ and $p_4$ come into play. 
We have a pair of conjugated solutions, which are given by:
\begin{eqnarray}
 \left\{\begin{matrix}
   \alpha_{1,2}&=&\mp \frac{1}{6} \sqrt{\frac{1}{3} \left(5 \sqrt{73}-23\right)};\\
    \alpha_{2,2}&=&\mp \frac{1}{108} \left(\sqrt{73}-5\right) \sqrt{15 \sqrt{73}-69}
    \end{matrix}\right.
\end{eqnarray}
and 
\begin{eqnarray}
 \left\{\begin{matrix}
   \alpha_{1,2}&=&\mp \frac{1}{6} i \sqrt{\frac{1}{3} \left(5 \sqrt{73}+23\right)};\\
    \alpha_{2,2}&=& \pm \frac{1}{108} i \left(\sqrt{73}+5\right) \sqrt{15 \sqrt{73}+69}
    \end{matrix}\right.
\end{eqnarray}

Note that these solutions are sufficient for the two-loop UV-finiteness. Indeed, let us consider the action of the ${\cal R}'$ operation (see for details and conventions  \cite{BogoliubovBook}) on the two-loop amplitude $A^{l=2}$. The result will be given by 
\begin{equation}
{\cal R}'~ A^{l=2}= \left(1-\sum_\gamma (K{\cal R}')_\gamma \right)  A^{l=2}.
\label{str}
\end{equation}
By definition ${\cal R}'~ A^{l=2}$ must be equal to a local expression. In our case $\sum_\gamma (K{\cal R}')_\gamma ~A^{l=2}$ can contain the $1/\epsilon^2$, $1/\epsilon$ and $\log/\epsilon$ terms.  However, if $\sum_\gamma (K{\cal R}')_\gamma  A^{l=2}$ is local, as well means that $A^{l=2}$ can contain only $1/\epsilon$ local poles too. One can see that actually 
\begin{equation}
\sum_\gamma (K{\cal R}')_\gamma ~ A^{l=2}\sim 
~\mbox{sing.}~\left(A_{d.t.}+ A_{s.t.}\right)\ldots
\label{str}
\end{equation}
and is equal to zero under the one-loop finiteness condition. That means that $A^{l=2}$ under the one-loop fineness condition can contain only the $1/\epsilon$ pole which in turn can be cancelled in the sum with one-loop result by adding higher orders of $g$ expansion into double-trace terms. 

Alternatively, one can consider $p_3^2=p_4^2=0$ and fix one-loop freedom in the fixed point solution by the requirement of non-local term cancellation. Also, as was stated before, the off-shell condition on external legs makes consideration of
the amplitude very similar to that of two-point correlation functions \cite{Grabner:2017pgm}, i.e. every diagram contributing to the amplitude with off-shell legs can be considered as a subgraph of the diagram contribution to the correlation function.

Direct computations in the two-loop case are rather complicated because one has to take into account $\sim300$ individual diagrams. The situation at higher loops is even more involved but using intuition from $D=4$, one can hope that a similar pattern of cancellation will hold in all orders of PT. Indeed, the colour structure, as well as the topology, of our $D=6$ diagrams
are identical to the $D=4$ counterparts; so we think that the general arguments of \cite{Pomoni:2008de,Dymarsky:2005uh} can be applied in the $D=6$ situation as well. 

\newpage
\bibliographystyle{unsrt}
\bibliography{refs_d6}

\end{document}